%% file: main.tex
\documentclass[reprint,amsmath,nofootinbib,showkeys,fleqn,amssymb,aps,prd, superscriptaddress]{revtex4-2}

\usepackage{newtxtext,newtxmath}
\usepackage[T1]{fontenc}
\usepackage[utf8]{inputenc}

\newcommand\farcs{\hbox{$.\!\!^{\prime\prime}$}} 

\DeclareRobustCommand{\VAN}[3]{#2}
\let\VANthebibliography\thebibliography
\def\thebibliography{\DeclareRobustCommand{\VAN}[3]{##3}\VANthebibliography}

\usepackage{graphicx}	
\usepackage{amsmath}	
\usepackage{orcidlink}
\usepackage{xcolor}
\definecolor{DarkGreen}{rgb}{0,0.6,0}

\newcommand{{\qsub}}{\bf{q}}
\newcommand{{\rsub}}{{\bf{r}}}
\newcommand{{\xlight}}{{\bf{n}}_{\rm{light}}}
\newcommand{{\xlens}}{{\bf{n}}_{\rm{mac}}}
\newcommand{{\data}}{\bf{D}}
\newcommand{{\datan}}{{\bf{d}}_{\rm{n}}}
\newcommand{{\datanprime}}{{{\bf{d}}_{\rm{n}}^{\prime}}}
\newcommand{{\dimg}}{{\bf{d}}_{\mathcal{I}}}
\newcommand{{\dimgprime}}{{\bf{d}}_{\mathcal{I}}^{\prime}}
\newcommand{{\dptsrc}}{{\bf{d}}_{\rm{ptsrc}}}
\newcommand{{\dptsrcprime}}{{\bf{d}}_{\rm{ptsrc}}^{\prime}}
\newcommand{{\dfr}}{{\bf{d}}_{\rm{fr}}}
\newcommand{{\dfrprime}}{{\bf{d}}_{\rm{fr}}^{\prime}}
\newcommand{{\zlens}}{{z_{\rm{d}}}}
\newcommand{{\zsrc}}{{z_{\rm{s}}}}

\def\sigsub{\Sigma_{\rm{sub}}}
\def\dlos{\delta_{\rm{los}}}
\def\mhm{m_{\rm{hm}}}
\def\sersic{S\'ersic}

\begin{document}

\title{JWST Lensed Quasar Dark Matter Survey III: Dark Matter Sensitive Flux Ratios and Warm Dark Matter Constraint from the Full Sample}

\author{R. E. Keeley$^\dagger$\orcidlink{0000-0002-0862-8789}}
 \affiliation{University of California, Merced, 5200 N Lake Road, Merced, CA 95341, USA}

\author{A.~M.~Nierenberg$^\dagger$}
\email{anierenberg@ucmerced.edu}
\affiliation{University of California, Merced, 5200 N Lake Road, Merced, CA 95341, USA}

\author{D.~Gilman}
\affiliation{Department of Astronomy $\&$ Astrophysics, University of Chicago, Chicago, IL 60637, USA}
\affiliation{Brinson Prize Fellow}

    \author{T.~Treu\orcidlink{0000-0002-8460-0390}}
	\affiliation{Department of Physics and Astronomy, University of California, Los Angeles, CA,  90095, USA}

  \author{X.~Du\orcidlink{0000-0003-0728-2533}}
    \affiliation{Department of Physics and Astronomy, University of California, Los Angeles, CA,  90095, USA}

\author{C.~Gannon\orcidlink{0009-0009-0443-3181}}
    \affiliation{University of California, Merced, 5200 N Lake Road, Merced, CA 95341, USA}

\author{P. ~Mozumdar \orcidlink{0000-0002-8593-7243}}
\affiliation{Department of Physics and Astronomy, University of California, Los Angeles, CA, 90095, USA}
\affiliation{Department of Physics and Astronomy, University of California, Davis, 1 Shields Ave., Davis, CA 95616, USA}

\author{K.~C.~Wong\orcidlink{0000-0002-8459-7793}}
    \affiliation{Research Center for the Early Universe, Graduate School of Science, The University of Tokyo, 7-3-1 Hongo, Bunkyo-ku, Tokyo 113-0033, Japan}

 \author{H.~Paugnat\orcidlink{0000-0002-2603-6031}}
    \affiliation{Department of Physics and Astronomy, University of California, Los Angeles, CA,  90095, USA}

    \author{S.~Birrer\orcidlink{0000-0003-3195-5507}}
    \affiliation{Department of Physics and Astronomy, Stony Brook University, Stony Brook, NY 11794, USA}

    \author{M.~Malkan\orcidlink{0000-0001-6919-1237}}
    \affiliation{Department of Physics and Astronomy, University of California, Los Angeles, CA,  90095, USA}

    \author{A.~J.~Benson\orcidlink{0000-0001-5501-6008}}
    \affiliation{Carnegie Institution for Science, Pasadena CA 91101, USA }

    \author{K.~N.~Abazajian\orcidlink{0000-0001-9919-6362}}
    \affiliation{Department of Physics and Astronomy, University of California, Irvine, CA 92697-4575, USA}
    
    \author{T.~Anguita\orcidlink{0000-0003-0930-5815}}
    \affiliation{Instituto de Astrofisica, Departamento de Ciencias Fisicas, Universidad Andres Bello, Chile}
    \affiliation{Millennium Institute of Astrophysics, Chile}
    
    \author{V.~N.~Bennert\orcidlink{0000-0003-2064-0518}}
    \affiliation{Physics Department, California Polytechnic State University, San Luis Obispo, CA 93407, USA }
    
    \author{S.~G.~Djorgovski\orcidlink{0000-0002-0603-3087}}
    \affiliation{California Institute of Technology, Pasadena CA 91125, USA }
    
    \author{S.~F.~Hoenig\orcidlink{0000-0003-3030-2360}}
    \affiliation{School of Physics and Astronomy, University of Southampton, Southampton SO17 1BJ, United Kingdom }
    
    \author{A.~Kusenko\orcidlink{0000-0002-8619-1260}}
    \affiliation{Department of Physics and Astronomy, University of California, Los Angeles, CA,  90095, USA}
    \affiliation{Kavli Institute for the Physics and Mathematics of the Universe (WPI), UTIAS, The University of Tokyo, Kashiwa, Chiba 277-8583, Japan}

    \author{H.~R.~Larsson\orcidlink{0000-0002-9417-1518}}
    \affiliation{University of California, Merced, 5200 N Lake Road, Merced, CA 95341, USA}

    \author{T.~Morishita\orcidlink{0000-0002-8512-1404}}
    \affiliation{IPAC, California Institute of Technology, MC 314-6, 1200 E. California Boulevard, Pasadena, CA 91125, USA}
    
    \author{V.~Motta\orcidlink{0000-0003-4446-7465}}
    \affiliation{Instituto de F\'{\i}sica y Astronom\'{\i}a, Universidad de Valpara\'{\i}so, Avda. Gran Breta\~na 1111, Valpara\'{\i}so, Chile }
    
    \author{L.~A.~Moustakas}
    \affiliation{Jet Propulsion Laboratory, California Institute of Technology, 4800 Oak Grove Dr, Pasadena, CA 91109}

    \author{W.~Sheu\orcidlink{0000-0003-1889-0227}}
    \affiliation{Department of Physics and Astronomy, University of California, Los Angeles, CA,  90095, USA}
    
    \author{D.~Sluse\orcidlink{}}
    \affiliation{STAR Institute, University of Li\`ege, Quartier Agora - All\'ee du six Ao\^ut, 19c B-4000 Li\`ege, Belgium }
    
    \author{D.~Stern\orcidlink{}}
    \affiliation{Jet Propulsion Laboratory, California Institute of Technology, 4800 Oak Grove Dr, Pasadena, CA 91109}

   \author{M.~Stiavelli\orcidlink{0000-0002-8512-1404}}
    \affiliation{Space Telescope Science Institute, 3700 San Martin Drive, Baltimore, MD 21218, USA}
    
    \author{D.~Williams\orcidlink{0000-0002-8386-0051}}
    \affiliation{Department of Physics and Astronomy, University of California, Los Angeles, CA,  90095, USA}

\begin{abstract}
We present the full sample of measurements of the warm dust emission of 31 strongly-lensed, multiply imaged quasars, observed with JWST MIRI multiband imaging, which we use to constrain the particle properties of dark matter. 
The strongly lensed warm dust region of quasars is compact and statistically sensitive to a population of dark matter halos down to masses of $10^6$ M$_\odot$. The high spatial resolution and infrared sensitivity of MIRI make it uniquely suited to measure multiply imaged warm dust emission from quasars and thus to infer the properties of low-mass dark halos. In this work, we present image fluxes as well as dark matter sensitive warm dust flux ratios inferred from spectral energy distribution fitting. We then use these flux ratios to test for the presence of a Warm Dark Matter turnover in the halo mass function. We use a forward modeling 
pipeline which explores dark matter parameters while also accounting for tidal stripping effects on subhalos, globular clusters, and complex deflector macromodels with $m=1, m=3, \text{ and } m=4$ elliptical multipole moments. Adopting a comparable prior on the projected density of substructure to previous analyses, the data presented here provide a factor of 2 improvement in sensitivity to a turnover in the halo mass function. Assuming subhalo abundance predicted by the semi-analytic model {\tt{galacticus}}, we infer with a Bayes factor of 10:1, a half-mode mass $m_{\rm{hm}} < 10^{7.8} M_{\odot}$. If instead we use a prior from $N$-body simulations, we infer $m_{\rm{hm}} < 10^{7.6} M_{\odot}$. This corresponds to a thermally produced dark matter mass of 5.6 (6.4) keV for the {\tt galacticus} and $N$-Body priors respectively. This is one of the strongest constraints to date on a turnover on the halo mass function, and the flux ratios and inference methodology presented here can be used to test a broad range of dark matter physics.

\end{abstract}


\maketitle

\def\thefootnote{$\dagger$}\footnotetext{These authors contributed equally to this work}\def\thefootnote{\arabic{footnote}}

\section{Introduction}

A major goal of modern physics is understanding the properties of dark matter, including whether it interacts with itself or with ordinary matter, and its formation history. Common probes of the nature of dark matter include direct detection, indirect detection, or production at colliders, all involving potential interactions with the Standard Model.  These methods have placed important limits on interaction cross sections between dark matter and standard model particles \cite{Bertone_dm_detection}; however, astrophysical gravitational probes remain the only confirmed source of detection of dark matter.

The mass function and internal density profiles of low mass dark matter halos (M$\lesssim10^{10}$ M$_\odot$) are particularly sensitive to the particle properties of dark matter and even the evolutionary history of the Universe~\citep{bode_halo_2001, schneider_non-linear_2012, bose_copernicus_2016, ludlow_mass-concentration-redshift_2016,bullock_small-scale_2017,buckley_gravitational_2018}. As the dark matter free-streaming length is increased from Cold Dark Matter (CDM) to Warm Dark Matter (WDM), low-mass dark matter halos can become less dense and less abundant~\citep{bode_halo_2001, schneider_non-linear_2012, bose_copernicus_2016, ludlow_mass-concentration-redshift_2016}, while self-interactions can drive a diverse range of observed present-day density profiles depending on the details of the interactions \citep{kamada_17,buckley_gravitational_2018, gilman_constraining_2023}.

Low-mass structure can be studied down to masses of $\sim 10^8$ M$_\odot$ in the Local Group using the faint galaxies that form within the dark matter halos~\citep{bullock_small-scale_2017, nadler_dark_2021}. Studies have used the abundance of these satellite galaxies to place strong limits on the free-streaming length of dark matter ~\citep{nadler_milky_2020} . 

One can also study low-mass halo properties without relying on internal galaxies for detection. One way to achieve this is by looking for the imprint of Local Group subhalos on stellar streams ~\citep{Bovy_stream,banik_stream_2018,Bonaca_stream,banik_evidence_2021,banik_evidence_2021_2}.  Outside of the Local Group, absorption from gas in low-mass halos along quasar sight lines (the Lyman-alpha forest) has been used to place limits on the abundance of low-mass halos \citep{Viel_2013_Lya,Irsic_2017_Lya,Villasenor_2023_Lya}, with results consistent with measurements of Milky Way satellite galaxies. 

Strong gravitational lensing provides an alternative and powerful approach to measuring the properties of low-mass dark matter halos. It is sensitive to low-mass halos outside of the Local Group, and does not require the dark matter halos to contain detectable baryons. In a galaxy-scale strong gravitational lens, a background light source is multiply imaged by a massive foreground galaxy. The image positions are determined by the first derivative of the gravitational potential and are therefore particularly sensitive to the large scale mass distribution governed by the main lens and its dark matter halo. The image magnifications depend on the second derivative of the gravitational potential and are therefore sensitive to perturbations by low-mass halos \citep[e.g.][and references therein]{treu10, vegetti_strong_2023}. 

Unresolved sources are extremely sensitive to low-mass perturbations given their small angular sizes. Such sources can be significantly magnified relative to the smooth mass distribution by halos with virial masses below $10^7$ M$_\odot$. There is a lower limit on the usable source size, however. Specifically, it is important that the source be larger than $\sim \mu$as angular ($\sim 1$~ pc physical) scales to avoid significant lensing by stars in the lens galaxy, a phenomenon known as optically thick microlensing \citep{Wambsganss_microlensing}. \citet{mao_evidence_1998}, \citet{metcalf_small-scale_2012}, and \citet{dalal_direct_2002} used quadruply imaged radio jets to reveal the presence of additional structure beyond the smooth mass distribution of the main lens. However, quadruply imaged radio jets are extremely rare:  only 7 were known at the time of those works. \citet{nierenberg_detection_2014, nierenberg_probing_2017, nierenberg_double_2020} demonstrated that quasar nuclear narrow-line emission \citep[as proposed by][]{moustakas_detecting_2003} could also be used to measure the properties of dark matter, significantly expanding the sample of lenses which could be used. The quasar warm dust is more compact ($\sim 1-10$~pc  \citep{burtscher_diversity_2013, leftley_parsec-scale_2019, honig_redefining_2019}) than the quasar nuclear narrow-line emission ($\sim 60-100$~pc \cite{Nierenberg_0405_2023,muller-sanchez_outflows_2011,nierenberg_detection_2014}), and therefore sensitive to lower-mass perturbations than the quasar narrow-line region \cite{Nierenberg_0405_2023} while still avoiding perturbations from microlensing.

In this series of papers we report measurements of lensed quasar emission, and dark matter constraints from the warm dust region using the Mid Infrared Instrument (MIRI) on JWST \citep{Rieke_MIRI_introduction}.  These data were obtained from the JWST lensed quasar dark matter survey (GO-2046, PI: Nierenberg).  In Paper I of this series, we presented the overall survey strategy and flux measurement methods \citep[][hereafter N24]{Nierenberg_0405_2023}. In Paper II \cite[][hereafter K24]{Keeley_2024}, we presented warm dust flux-ratio measurements from the first 9 systems which were observed in this program and used these data to place a new constraint on the half-mode mass of warm dark matter models. In this work, we report warm dust flux ratios for the full sample of 31 lenses, and use these data in combination with narrow-line flux ratio measurements and improved analysis methods that we discuss in detail in a companion paper - Paper IV in this series \citet[][hereafter G25]{JWST_4_Gilman}, to place new constraints on warm dark matter models.

This paper is organized as follows. In Sec.~\ref{sec:obs}, we describe the observation methods and initial data reduction. In Sec.~\ref{sec:fluxes}, we describe how we fit the multi-band imaging data to extract the point source fluxes, and report these results in Sec.~\ref{sec:lens_modeling_results}. In Sec.~\ref{sec:SED}, we describe how we fit the multi-band point source fluxes with a spectral energy distribution model to extract the flux ratios of the warm dust component, and report these results in Sec.~\ref{sec:SED_results}. In Sec.~\ref{sec:WDM} we describe the components of the warm dark matter model and the inference methodology. We present the posterior of the warm dark matter model in Sec.~\ref{sec:WDM_results}. We discuss our findings and summarize our key results in Sec.~\ref{sec:conclusions}.

We note that while this paper summarizes key aspects of the structure formation model as well as the lens modeling approach used in the dark matter inference, more detail can be found in a companion publication, G25 which describes these methods in more depth. G25 also extends the dark matter inference relative to this work by including information from the full imaging data and lensed quasar host galaxy, in addition to the quasar point source positions and flux ratios which are used in this work.


\section{Observations and Reduction}\label{sec:obs}

Targets for the JWST lensed quasar dark matter survey were selected from known quadruply imaged quasars at the time of survey design (2021). In addition to being quadruply imaged, the lenses were selected to meet two additional criteria: i) the deflector must not be a disk galaxy based on available imaging; ii) the predicted faintest lensed image flux must be greater than 1 mJy based on unresolved WISE W4 (22 $\mu$m) imaging for lenses with source redshifts greater than 2, or a minimum total unresolved WISE W4 flux of 3 mJy for sources with redshifts less than 2. Criterion i) ensured that a robust lens model for the main deflector ("macromodel") could be obtained, while criterion ii) ensured a significant detection with MIRI. More details of the selection process can be found in N24. After these selections, we were left with a final sample of 31 lenses. Key properties of these lenses are listed in Table~\ref{tab:lens_sample}. 

Each lens was observed in four MIRI filters: F560W, F1280W, F1800W and either F2100W or F2550W. These filters have central wavelengths of 5.6, 12.8, 18.0 and 21.0 $\mu$m respectively. F2100W was used if the faintest image was predicted to be fainter than 1 mJy in WISE W4 and at a redshift lower than 2. The four filters were selected to enable simple modeling of the spectral energy distribution, in order to remove contamination from the quasar accretion disk and `hot' ($\sim$1500 K) dust associated with the dust sublimation radius, which may be microlensed depending on the quasar luminosity. Typical exposure times were $\sim$58 s in each of the three bluer filters and $\sim$600 s in the reddest filter. Observations in each filter were made using a three-point dither pattern to improve spatial resolution and mitigate the impact of cosmic ray hits.

\begin{table*}
    \centering
    \begin{tabular}{llllllll}
    \hline
    \hline
     Lens           &Abbrev. Name. & Source z & Lens z$^{a}$ & q$^\star$ $^b$ &    Obs. Date. & Discovery paper(s) & Reason for Exclusion \\
    \hline
     PS  J014710.1$+$453044 &PSJ0147     &   2.38          &    0.678 \citep{Goicoechea_0147_deflector_z} &  0.9 \cite{schmidt_strides_2023} & Aug. 11 23                  & \cite{Berghea_0147_discovery}   &                \\
     
     SDSS J024848.7$+$191331 &J0248     &   2.43          & 0.5$^*$  
     & 0.44 \cite{schmidt_strides_2023} & Aug. 11 23            &   \cite{shajib_is_2019, delchambre_gaia_2019}                    \\
     
     WISE J025942.9$-$163543  &J0259    & 2.16       & 0.904       & 0.44 \citep{shajib_is_2019}   & Jul. 30 23            & \citep{Schechter_0259_discovery}      &                     \\
     
     DES J040559.7$-$330851 &J0405    &1.713       &0.5$^*$  &0.74 \cite{schmidt_strides_2023} & Oct. 12 22    & \cite{anguita_strong_2018} & \\
     
     MG J041437.7+053443   & MG0414  & 2.64        &  0.96\cite{0414_spec_redshift}  &0.75 \cite{sluse_microlensing_2012}  &  Aug. 29 23   &  \cite{0414_discovery}       \\
     
     J043814.9$-$121714  &HE0435   &   1.69       &    0.46  &0.88 \cite{sluse_microlensing_2012}    &  Sep. 5 23  &   \cite{0435_discovery}       \\
     
     J045723.6$-$782048    & J0457   &  3.15           & -  &-           &  Jul. 7 23   &  \cite{lemon2023}            & Triple           \\
     GRAL J060710.8$-$215217 &J0607    &1.302       &0.55$^d$
     &0.75    & Feb. 22 23   & \cite{Stern_2021_gral, lemon2023}& \\
    
     GRAL J060841.4+422937 &J0608     & 2.345      &         0.5$^*$ & 0.5     & Feb. 23 23  &\cite{Stern_2021_gral, lemon2023}& \\
     
     GRAL J065904.1+162909 &J0659     & 3.083      & 0.766
      &0.95 \cite{schmidt_strides_2023}   & Feb. 27 23  &\cite{delchambre_gaia_2019, lemon2023} & \\
     
     J080357.7$+$390823 &J0803 &   2.97  &  0.5$^*$     &   0.8        & Nov. 6 23  & \cite{lemon2023} \\
     
     SDSS J092455.8$-$215217& J0924  & 1.52    & 0.39\cite{7_spec_redshifts}  & 0.83 \cite{sluse_microlensing_2012}    & Nov. 16 23   &  \cite{0924_discovery}\\
     
     W2M J104222.1+164115  &J1042     & 2.517      & 0.5985          & 0.75 \cite{schmidt_strides_2023} & Dec. 15 22 &\cite{Glikman_1042_2023} & \\
     
     J111623.5$-$065739& HE1113    & 1.24          &  0.75     &-     &   May 4 23   &  \cite{1113_discovery}  &  Deflector not detected \\
     
     J111816.9$-$074558 & PG1115    &  1.74    &     0.311\cite{1115_spec_redshift} &0.95 \cite{sluse_microlensing_2012}  & May 4 23   & \cite{1115_discovery} \\
     
     GRAL J113100.0$-$442000 & GRAL1131 & 1.09      &  0.472$^e$      &    0.6      &   May 27 23  &  \cite{Krone_Martins_Gral_18} \\
     
     RX   J113151.5$-$123158  & RXJ1131 &    0.659    &   0.258   & 0.9 \cite{Suyu_lensq_2013}    &   Jun. 24 23  &  \cite{Sluse_RXJ1131}  \\
     
     2M J113440.5$-$210322 & 2M1134     &  2.77         & 0.66     &  0.74\cite{schmidt_strides_2023}                &  May 27 23      & \cite{2m1134_discovery}   \\
     
     SDSS J125107.6+293540 &J1251 & 0.802      &   0.410\cite{1251_spec_redshift}   & 0.67 \cite{schmidt_strides_2023} & May 3 23  &  \cite{1251_discovery}    \\
     
     J141546.2$+$112943    &H1413       & 2.56    & 0.5$^*$   &   $^f$         &  Jul. 7 23     &  \cite{1413_discovery}    \\

     J153725.3$-$301017  &J1537       & 1.721   &  0.592   & 0.76 \cite{schmidt_strides_2023} & Mar. 7 23  &\cite{lemon_gravitationally_2018} \cite{delchambre_gaia_2019} \\
     
     PS J160600.2$-$233322    &J1606        & 1.696      & 0.92$^g$
     & 0.58 \cite{schmidt_strides_2023}      &  Mar. 8 23   & \cite{lemon_gravitationally_2018} \\
     
     J201749.1$+$620443   &J2017        & 1.72     &  0.201   & -           &  Jul. 7 23   & \cite{Stern_2021_gral}   & Spiral galaxy deflector    \\
     
     WFI J202610.4$-$453627    &J2026        & 2.23       &  0.5$^*$    &  0.9 \cite{Chantry_2010}            & Apr. 15 23  & \cite{Morgan_2026_2033_2004} \\
     
     WFI J203342.2$-$472344  & WFI2033 &  1.66    &  0.66\citep{Eigenbrod_8lenses}       &   0.82 \cite{sluse_microlensing_2012}           & May 4 23 &  \cite{Morgan_2026_2033_2004}     \\
     
     DES J203802.7$-$400814 &J2038    & 1.19      & 0.230   &0.63 \cite{schmidt_strides_2023}     & April 18 23  & \cite{agnello_meets_2018} \\
     
     J204720.4+264401       & B2045 & 2.35  &  0.87  & -  & June 2 23    &   \cite{2045_discovery}    & High image magnification  \\
     
     J210752.4$-$161131       &J2107   &    2.67     &    -          & -  & May 4 23        &  \cite{2107_discovery}   &Doubly imaged in MIRI    \\
     
     J214505.1$+$634541 &J2145  &     1.56    &  0.5$^*$             &  0.72 \cite{schmidt_strides_2023} & Aug. 1 23      &   \cite{Lemon_GraL_2145_discovery}   \\
     
     J220544.2$-$372701 &J2205 &    1.85   &  0.63$^h$       &0.8 \cite{schmidt_strides_2023}      &  Jul. 1 23   &  \cite{lemon2023}          \\
     
     WISE J234416.9$-$302526  & J2344  &  1.3 &  0.5$^*$       & 0.84 \cite{schmidt_strides_2023} &  Jul. 11 23   &  \cite{2344_discovery}    \\
     
    \end{tabular}
    \caption{Information about the lens systems and observation details. ($a$) References are provided for deflector redshift measurements when the reference is different from the discovery paper. 
    ($b$) The axis ratio of the deflector in optical HST imaging, used as a prior in the dark matter lens modeling as described in \ref{sec:lens_modeling}. (
    $c$) Deflector redshift is set to 0.5 when a spectroscopic redshift has not been measured.  
    ($d$) (this paper; Appendix~\ref{app:spectra}), 
    ($e$) Sluse et al. (2025, in prep). 
    ($f$) Because the axis ratio for the deflector of H1413 is not well measured, we adopt a uniform prior for this parameter during lens modeling for this system.
    ($g$) Langeroodi et al. (2025, in prep.) 
    ($h$) K.C. Wong et al. (2025, in prep).} 
    \label{tab:lens_sample}
\end{table*}

Following N24 and K24, the initial observation reduction was done with the default JWST pipeline. Lenses published in K24 were reduced using CRDS version 11.16.21, while lenses analyzed in this paper were reduced with an updated version of the pipeline, CRDS 11.17.20.  Estimates of key JWST detector and optical parameters continue to be updated, however, we expect these updates to have minimal impact on the measured warm dust flux ratios which we use for our dark matter inference, because we assume $\sim$10\% absolute flux uncertainties, which is much larger than the estimated systematic flux uncertainty of the instrument of $\sim 3\%$ \citep{Gordon_MIRI_Calibration_25}.  Backgrounds were estimated and subtracted prior to drizzling onto the native pixel scale of 0\farcs11, using a customized routine\footnote{\url{https://github.com/STScI-MIRI/Imaging_ExampleNB}}.

In addition to the quasar flux ratio measurements presented in the next section, in Appendix \ref{app:spectra} we present redshift measurements for two systems, J0607-2152 and J2017+6204. 

\section{Quasar flux measurements}\label{sec:fluxes}
We measured the multi-band quasar point source fluxes following the  procedure that was used by K24. Here we provide a brief summary of the method.

The lensed quasar images appear as point sources in MIRI imaging and can be modeled by the point spread function (PSF) of the instrument. In the majority of lenses the lensed quasar host galaxy and lens galaxy were also apparent, typically in the bluest filters (F560W and F1280W). The lensed quasar fluxes were measured by adopting a forward modeling procedure in which we simultaneously fit the PSF, a gravitational lens model for the extended quasar host galaxy emission, and a light distribution model for the quasar host galaxy and lens galaxy, if observed in a given filter. 

The quasar host galaxy was modeled as a circular S\'ersic \cite{sersic_influence_1963} light distribution. Some quasar hosts required additional complexity. For these we added shapelets in addition to the circular \sersic~source to provide an adequate fit to the data. Typically, this complexity was only required in the bluer filters, with the notable exception of RXJ1131 which required much higher order shapelets than the other lenses and in all four filters. RXJ1131 is exceptional for its very low source redshift of 0.66. This enables the detection of low surface brightness fluctuations not detectable in other quasar host galaxies given the exposure depth. In addition to shapelets, for RXJ1131, we use two \sersic~light components offset from the main light source. Previous studies of RXJ1131 have also inferred a complex, multi-component source model \citep{Suyu_lensq_2013, Birrer_1131}. 

 Lens galaxies were modeled as elliptical \sersic~profiles with \sersic ~ index  fixed to 4. We kept the indices fixed as we found that they were typically poorly constrained given the depth of our observations and the brightness of the quasar images relative to the lens galaxy. 

Forward modeling was performed using {\tt lenstronomy} \citep{birrer_lenstronomy_2018, birrer_lenstronomy_2021}\footnote{Version 1.11.2}. When the lensed quasar host galaxy was detected, a lens model was included. The lens model was assumed to be a power-law elliptical mass distribution (PEMD) with variable density slope, plus external shear. If satellite galaxies were detected close in projection to the lens, they were modeled as singular isothermal spheres. The initial lens model used the quasar image positions as a constraint, however, in the final step of the analysis, this constraint was relaxed, and the quasar images were treated as independent point sources in the image plane. This final step reduced the dependence of the measured fluxes on the underlying lens model.

The PSF was modeled using {\tt webbpsf} \citep{perrin_simulating_2012, perrin_updated_2014-1}\footnote{Version 1.1.2.dev111+ga20ae48}, with temperature and `jitter' optimized as a part of the fitting routine. In F560W, we also included the cruciform artifact ~\citep{gaspar_quantum_2021, wright_mid-infrared_2023}. Because its strength varied significantly between targets, we added an additional parameter to the model, $f$, defined as the fractional weight of the cruciform artifact for a given observation.

\section{Results of Image Fitting}
\label{sec:lens_modeling_results}

\input{image_fitting_results}

\section{Spectral Energy Distribution Fitting}\label{sec:SED}
\input{sed_fitting_method}

\section{Results of spectral energy distribution fitting}\label{sec:SED_results}
\input{sed_fitting_results}


\section{Warm Dark Matter Inference Methods}\label{sec:WDM}
The new flux ratios presented in this paper will enable tests of any dark matter model that alters the properties of low-mass halos. In this first analysis with the complete dataset, we test  the same warm dark matter (WDM) model considered in previous flux-ratio analyses by \citet{gilman_warm_2020} and K24. 

Relative to previous work, this analysis implements several improvements in the methodology related to both the calculation of a likelihood function and the structure formation model. We refer to a companion paper, G25 for an in-depth discussion regarding the new lens modeling methodology, improved structure formation model, including the model for subhalo tidal evolution and free-streaming effects.
 This section gives a high-level overview of the essential features of the model and procedures.   

In Subsection \ref{sec:inference_framework}, we begin by stating the Bayesian inference problem we solve. We then review the dark matter model, and introduce the parameters to be measured with the data. Further discussion related to the Bayesian inference framework, the dark matter and subhalo tidal evolution model, and lens modeling methodology can be found in Sections II, IV, and IV of G25, respectively. 

\subsection{Bayesian inference framework}
\label{sec:inference_framework}
Our goal is to compute the posterior distribution: 
\begin{equation}
    \label{eqn:bayestheorem}
    p \left(\qsub | \data \right) \propto \pi \left(\qsub \right) \prod_{n=1}^{N} \mathcal{L}\left(\datan | \qsub \right),
\end{equation}
where $\boldsymbol{q}$ is a set of hyper-parameters describing a dark matter model, $\data$ is the complete dataset, $\pi\left(\qsub\right)$ is a prior probability, and $\datan$ represents the observations of the $n$th lens system. For each lens, we have measured the image positions $\dptsrc$ and flux ratios $\dfr$, such that $\datan = \left(\dptsrc, \dfr\right)$. The likelihood function is given by: 
\begin{eqnarray}
    \label{eqn:likelihood}
    \nonumber  \mathcal{L}\left(\dptsrc, \dfr | \qsub \right) =  \int p\left(\dptsrc, \dfr | \rsub, \xlens \right) \\
    \times p\left(\rsub | \qsub \right) p\left(\xlens\right) d \xlens d \rsub,
\end{eqnarray}
The first term represents the probability of measuring $\dptsrc$ and $\dfr$, given a set of macromodel parameters $\xlens$ and a {\textit{realization}} of halos, $\rsub$. A realization refers to a set of parameters that specify the masses, positions, concentrations, redshifts, etc for a full population of dark matter subhalos, line-of-sight halos, and globular clusters. Relative to previous work, the inclusion of globular clusters is a new addition to the structure formation model. We refer to Section IV of G25 for additional details. The term $p\left(\rsub | \qsub \right)$ encodes the probability of having a realization $\rsub$ in a dark matter model $\qsub$. Finally, $p\left(\xlens\right)$ represents a prior on the macromodel parameters.  We note that in this work, we only use the image positions and flux ratios to constrain the likelihood, but improved constraints can be obtained when including full information the imaging data. In particular, information from the lensed quasar host galaxy significantly reduces the uncertainty on model predicted flux ratios \citep[e.g.][]{2024GilmanTurbo}. We present the methodology for this approach and use this additional data when performing the dark matter inference in G25.

We evaluate Equation \ref{eqn:likelihood} using a forward modeling inference framework in which we reconstruct an observed lens system millions of times in the presence of different realizations. We generate realizations of WDM halo populations using the open-source code {\tt{pyHalo}}\footnote{\url{https://github.com/dangilman/pyHalo}}\citep{gilman_warm_2020}, which accounts for the second term in the integrand of Equation \ref{eqn:likelihood}, $p\left(\rsub | \qsub\right)$. The first term in the integrand, $p\left(\dptsrc, \dfr | \rsub, \xlens \right)$, involves generating model-predicted image positions and flux ratios given $\xlens$ and $\rsub$. This task requires lens modeling. We perform lensing calculations with {\tt{lenstronomy}} \citep{birrer_lenstronomy_2018,birrer_lenstronomy_2021}, and use the open-source code {\tt{samana}}\footnote{\url{https://github.com/dangilman/samana}} to wrap {\tt{lenstronomy}} and {\tt{pyHalo}} routines to generate millions of different lens models per lens system in order to evaluate the integral. 

The probability $p\left(\dptsrc, \dfr | \rsub, \xlens \right)$ involves an astrometric and a flux-ratio likelihood. Following the approach presented by \citet{gilman_probing_2019}, we handle the astrometric likelihood by reconstructing each lens with a different draw of statistical measurement uncertainties applied to the observed image positions, and then solve for a subset of macromodel parameters such that the lens equation is satisfied for these perturbed image coordinates. 

In previous analyses, we approximated the flux ratio likelihood using Approximate Bayesian Computing (ABC). In this work, we use an explicit flux-ratio likelihood that we evaluate as a multi-variate Gaussian. The covariance matrix is computed from the SED fitting, as discussed in Section \ref{sec:SED}. The motivation for using ABC in previous analyses were twofold. First, previous measurements of narrow-line emission using HST data \citep{nierenberg_double_2020} quoted uncertainties on image fluxes, which propagate non-linearly onto the flux ratios, and ABC accounts for non-Gaussian uncertainties in the forward model. In this work, the flux ratio uncertainties are well-approximated by a multi-variate Gaussian, which obviates this first motivation for using ABC. Second, it is unlikely that any given lens model will match the observed flux ratios. ABC allows for a relaxed tolerance threshold when accepting or rejecting samples, allowing one to approximate the likelihood function from a limited number of realizations. We have overcome this limitation by investing orders of magnitude more computational resources to obtain more model-predicted datasets. The total number of realizations generated for this analysis, and the analysis presented by G25, exceeds 174 million. 

Throughout the rest of this section, we discuss the important features of the dark matter model we constrain with data (Section \ref{sec:dark_matter_model}), and review the parameterization of the macromodel, including which parameters we sample from $p\left(\xlens\right)$, and which parameters are optimized to solve the lens equation for each realization (Section \ref{sec:lens_modeling}). For additional details regarding these topics, including the implementation of an improved tidal evolution model, and the modeling of globular clusters, we refer the reader to Section IV of G25. 

\subsection{Dark matter modeling}
\label{sec:dark_matter_model}

The first step in our modeling process is the generation of a realization of a population of dark matter halos in the lens (subhalos) and along the line-of-sight from the source to the observer (field halos). The properties of these halos depend on the details of the dark matter model, as well as the lens galaxy mass and redshift. We describe the specifics of these models below.

\subsubsection{Field Halos}
We parametrize the mass function for halos along the line-of-sight as
\begin{equation}
\frac{d^2N}{dM dV} = \delta_{\rm{LOS}}\left(1+\xi_{\rm 2halo}\left(M_{\rm{host}},z_d\right)\right)\frac{d^2N}{dM dV} \bigg |_{\rm{ST}},
\end{equation}
where $M_{\rm{host}}$ and $z_d$ are the mass and redshift of the host halo, and $\frac{d^2N}{dM dV}|_{\rm ST}$ is the Sheth-Tormen mass function~\cite{Sheth_Torman_2001}. The term $\xi_{2\rm{halo}}$ accounts for correlated structure around the main deflector, as described by \citet{gilman_probing_2019}. In our calculation of $\xi_{\rm{2halo}}$, we also include the modification calibrated against N-body simulations by \citet{Lazar++21}, which increases the number of halos introduced through this term by a factor of $\sim 2$.  The parameter $\delta_{\rm LOS}$ scales the entire halo mass function and is introduced to account for theoretical uncertainties in calculating the normalization of the halo mass function~\citep{2016MNRAS.456.2486D}.

In this work we test for the presence of a turnover in the halo mass function, associated with WDM models with a finite free-streaming length. Such a turnover is calculated as a suppression applied to the Sheth-Torman mass function, with the following fitting formula:
\begin{equation}
    \frac{d^2N_{\rm WDM}}{dM dV} = \frac{d^2N_{\rm CDM}}{dM dV} f_{{\rm{WDM}}}\left(M, m_{\rm{hm}}\right),
\end{equation}
where 
\begin{equation}
\label{eqn:mfuncsuppression}
f_{{\rm{WDM}}}\left(M, m_{\rm{hm}}\right) = \left(1 + a\left(\frac{M_{\rm hm}}{M}\right)^{b}\right)^{c},
\end{equation}
with $a = 1.95$, $b = 0.8$, and $c = -1.0$ \citep{Lovell+2020}. 

We model the internal density profile of halos as truncated Navarro-Frenk-White (NFW) profiles: \citep{navarro_universal_1997}
\begin{equation}
\label{eqn:densityprof}
\rho\left(r\right) = \frac{\rho_s}{x \left(1+x\right)^2}\frac{\tau^2}{x^2 + \tau^2},
\end{equation}
where $x \equiv r / r_s$, $\tau \equiv r_t / r_s$, $r_s$ is the halo scale radius, and $r_t$ is a truncation radius. We choose a truncation radius $r_t = r_{\rm{50}}$, where $r_{50}$ is the radius that encloses 50 times the critical density. This keeps the mass rendered along the line of sight finite while still preserving the total mass of the halos.

We calculate $r_s$ for a halo of mass $M$ using the mass-concentration relations modified for WDM expectations. For WDM models, the concentration of dark matter halos is lowered relative to the CDM expectation due to the delayed formation time of structure ~\cite{2002ApJ...568...52W,ludlow_mass-concentration-redshift_2014}.  We use the CDM mass-concentration relation presented by \citet{Diemer_2019_CDM_concentration}, modified to include the effects of WDM using the fitting function given by \citet{bose_copernicus_2016}:
\begin{equation}
\label{eqn:concentrationwdm}
    \frac{c_{\rm WDM}(M,z)}{c_{\rm CDM}(M,z)} = f_{\rm{WDM}}\left(M, m_{\rm{hm}}\right),
\end{equation}
where $f_{\rm{WDM}}$ is defined in Equation \ref{eqn:mfuncsuppression} in relation to the mass function, but the coefficients change to $a=1.0$, $b=60$, and $c=-0.17$ for the suppression to the concentration-mass relation.

\subsubsection{Subhalos}
Subhalos are dark matter halos that have accreted onto the host halo of the main deflector and are thus modified by the tidal field of the host. The projected number density of subhalos near the lensed quasar images depends on the mass and redshift of the main host as well as the details of tidal stripping. We describe how we incorporate these effects here.

We follow our previous work \citep{gilman_warm_2020} in describing the subhalo mass function as: 
\begin{equation}
\label{eqn:shmf}
    \frac{d^2N_{\rm sub}}{dM dA} = \frac{\Sigma_{\rm sub}}{10^8 M_\odot} \left(\frac{M}{10^8 M_\odot}\right)^\alpha \mathcal{F}(M_{\rm{host}},z_{\rm{d}}) f_{{\rm{WDM}}}\left(M, m_{\rm{hm}}\right).
\end{equation}

The subhalo mass function describes the properties of subhalos at the moment of infall, prior to the effects of tidal stripping. We therefore use the same coefficients $a=1.95$, $b=0.8$, and $c=-1.0$ as used in the field halo mass function suppression (Equation \ref{eqn:mfuncsuppression}). Here, $\Sigma_{\rm sub}$ and $\alpha$ are the normalization and logarithmic slope, while the term $\mathcal{F}$ captures the dependence of the normalization on the properties of the host galaxy mass and redshift. To compute the evolved subhalo mass function, which accounts for the tidal evolution of subhalos in the host, we use the tidal evolution model presented by \citet{Du++25}. We direct readers to Section 4 of G25 for further details. 

We consider a uniform prior for the slope $\alpha$ in the range $[-1.95,-1.85]$. This range is broad enough to encompass both the dark matter only expectation, as well as potential modification to the slope of the subhalo mass function by baryons \citep{Springel++08,fiacconi_cold_2016, Benson_2020_normalization}. The normalization $\Sigma_{\rm sub}$ is a major source of theoretical uncertainty. Current state of the art simulations differ by a factor of 2 \citep{Gannon_2025}. For our base simulations, we adopt a broad, uniform prior for $\log_{10} \Sigma_{\rm{sub}}$ in the range $[-2.2, 0.2]$. This extends a factor of ten above and below the prediction from N-body simulations $\log_{10} \Sigma_{\rm{sub}} \sim -1.3$ and semi-analytic models $\log_{10} \Sigma_{\rm{sub}} \sim -1.0$ \citep{Gannon_2025}. 

The function $\mathcal{F}$ captures the dependence of the normalization of the infall subhalo mass function on host halo mass and redshift \citep{gilman_warm_2020, Gannon_2025}:

\begin{equation}\label{eq:mathcalF}
\log\mathcal{F}(M_{\rm{host}},z_d) = k_1 \log \left(\frac{M_{\rm{host}}}{10^{13}M_{\odot}}\right) + k_2 \log \left(z + 0.5\right),
\end{equation}

\noindent 
where we adopt values of $k_1 =0.55$ and $k_2 = 0.37$, based on the results of \citet{Gannon_2025}.

Table \ref{tab:tableqsub} provides a summary of the dark matter parameters and priors used in this analysis.

	\begin{table*}
		\setlength{\tabcolsep}{12pt}
		\caption{\label{tab:tableqsub} Description of the dark matter hyper-parameters introduced in Section \ref{sec:dark_matter_model}. The predictions for $\delta_{\rm{LOS}}$, $\alpha$ and $\Sigma_{\rm{sub}}$ remain unchanged in WDM, as these parameters determine the form of the subhalo mass function anchored at high masses.}
		\begin{tabular}{cccc}
			\hline
			Hyper-parameter & Description & Sampling distribution &  Remarks\\	
			\hline
			$\delta_{\rm{LOS}}$ & rescales the amplitude of the& $\mathcal{U}\left(0.9, 1.1\right)$ & $\delta_{\rm{LOS}}=1$ corresponds to the \\
			& field halo mass function  & & Sheth-Tormen prediction\\ \\ 
			$\alpha$ & logarithmic slope of the & $\mathcal{U}\left(-1.95, -1.85\right)$ & CDM predicts $\alpha \sim -1.9$\\
			& subhalo mass function at infall & & \\ \\ 
			$\Sigma_{\rm{sub}} \left[\rm{kpc^{-2}}\right]$ & amplitude of the differential & $\log_{10} \mathcal{U}\left(-2.2, 0.2\right)$ & N-body predicts $\sim 0.05 \ \rm{kpc^{-2}}$\\
			& subhalo mass function at infall & & \tt{galacticus} predicts $\sim 0.1 \ \rm{kpc^{-2}}$ \\ \\ 
			$m_{\rm{hm}} \left[M_{\odot}\right]$ & half-mode mass scale & $\log_{10}\mathcal{U}\left(4.0, 10\right)$ & mass function $\&$ concentrations\\
			& & & suppressed for $ m \lesssim m_{\rm{hm}}$ \\
			\hline
		\end{tabular}
	\end{table*}

\subsection{Gravitational lens modeling}
\label{sec:lens_modeling}
Section V of G25 gives a detailed overview of the lens modeling considerations for each of the 26 lenses with warm dust measurements plus 2 systems with narrow-line flux ratio measurements used in this analysis. In this section, we review the essential features of the lens modeling assumptions. 

We model the main lens as a power-law elliptical mass distribution, plus external shear, and elliptical multipoles of order $m=1$, $m=3$, and $m=4$ \citep{Paugnat_elliptical_2025}. Luminous galaxies observed close in projection to the lensed images are included directly in the lens model as singular isothermal spheres (SIS). We note that the lens modeling done in this step is completely independent of the lens modeling of the MIRI images used to extract the quasar point source fluxes.  That lens modeling was used to enable measurement of the quasar image fluxes in the presence of the qso host galaxy light in a way that minimized dependence on the lens model, thus the point sources in those models were treated as independent foreground light sources.

As discussed in Section \ref{sec:inference_framework}, we solve for a subset of macromodel parameters such that the lens equation is satisfied for each realization of halos. In particular, we solve for the Einstein radius, mass centroid, ellipticity, position angle, external shear strength, and external shear angle. The mass axis ratio $q$, logarithmic profile slope of the main deflector $\gamma$, the multipole strengths and orientations $\left(a_1, a_3, a_4, \varphi_{1}, \varphi_{3}, \varphi_4\right)$, and the mass $\theta_{\rm{E,G2}}$ and position of a satellite galaxy (if present) are held fixed to values drawn from $p\left(\xlens\right)$. We draw $\gamma$ from a prior $\mathcal{N}\left(2.1, 0.1\right)$ based on measurements of galaxy scale lenses from the Sloan Lens ACS Survey \cite{Barnabe_gamma_slope}.

We draw $q$ from a truncated normal distribution centered on the observed ellipticity of the light of each deflector, $q^{\star}$, with a standard deviation of 0.2 and a lower bound of $q^{\star} - 0.2$. This prior is derived from measurements of the relationship between deflector light and mass from HST imaging by \citet{shajib_is_2019} and \cite{schmidt_strides_2023}, and enforces that the mass should not be significantly more elongated than the light. The axis ratios of the deflector light are derived from observations with HST imaging when available and MIRI imaging if HST imaging is not available. Table \ref{tab:lens_sample} lists the axis ratio value of the light and reference if not from this work. 

Multipole amplitudes are drawn from Gaussian priors ($a_1\sim\mathcal{N}(0, 0.005)$, $a_3\sim\mathcal{N}(0, 0.005)$, and $a_4\sim\mathcal{N}(0, 0.01)$) based on observations of massive elliptical galaxy isophotes. The $m=3$ and $m=4$ amplitude and direction priors are derived from \citet{hao_isophotal_2006} and  \citep{Oh_improving_2024}, and the prior for the $m=1$ amplitude and direction is from \citet{Amvrosiadis_m1_25}. In a typical galaxy-scale strong gravitational lens, baryons comprise approximately 2/3 of projected matter within the Einstein radius \citep{Auger_slacs_fdm}. By applying priors based on the baryonic light distribution, we are conservatively assuming the maximum possible effect of deviations from ellipticity on our lenses relative to the combined dark matter plus baryon projected mass distribution. 

Luminous satellites or companion galaxies, if present, are placed at their observed position in the image plane with an astrometric uncertainty of 50 milli-arcseconds. If a redshift has been measured for the object (as is the case for WFI2033 and HE0435), then we include the SIS at the measured redshift. Otherwise, the SIS is assumed to be at the redshift of the main deflector. Priors on the companion's Einstein radius $\theta_{\rm E, G2}$ are derived from a velocity dispersion measurement, if one is available. If one is not available, we use the Faber Jackson relation \citep{Faber_Jackson_76} to relate the ratio of the total flux of the satellite ($L_{\rm G2}$) to the total flux of the main lens ($L_{\rm G1}$)  using $\theta_{\rm E,G2} = \theta_{\rm E} \sqrt{L_{\rm G2}/L_{\rm G1}}$, where $\theta_{\rm E}$ is the Einstein radius of the main lens. This relies on the assumption that $\theta_{\rm E} \propto \sigma^2$ and the Faber Jackson relation: $L\propto \sigma^4$.

\section{Results}\label{sec:WDM_results}

This section presents the main results of our analysis. In Section \ref{sec:WDM_results} we present the constraints on the free-streaming length of dark matter obtained for the full sample of 28 lensed quasars. In Section  \ref{ssec:correction} we present a test of our SED fitting procedure by presenting the degree of inferred microlensing versus the strength of inferred millilensing. This test illustrates that our SED correction to the flux ratios removes microlensing contamination from the warm dust flux ratios, validating the motivation for targeting the 31 systems in the JWST sample with MIRI.

\subsection{Inference on WDM}
\begin{figure*}
    \includegraphics[width=\textwidth]{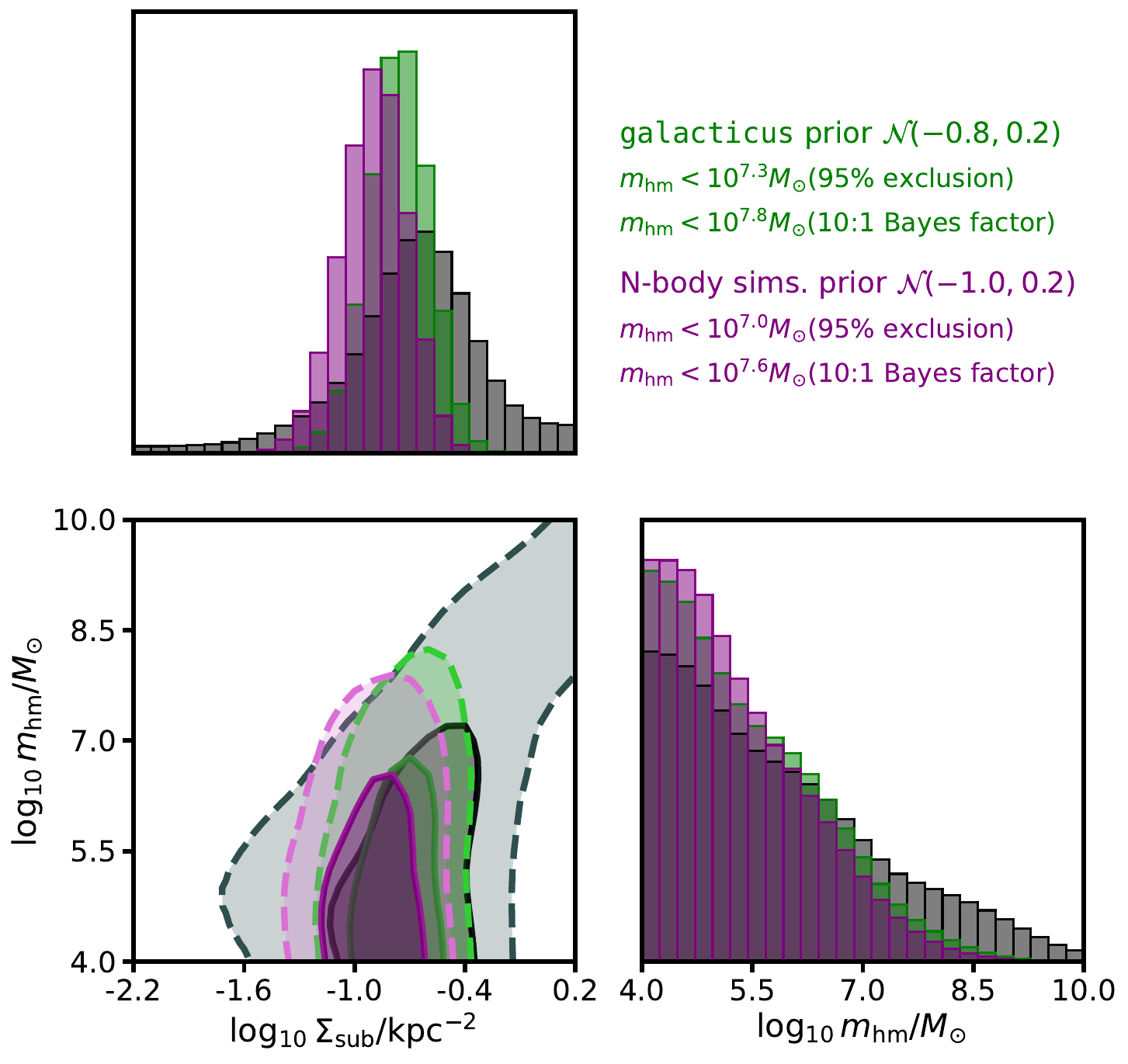}
    \caption{The inferred joint posterior probability distribution for the dark matter parameters for the full sample of 26 mid-IR plus 2 narrow-line lenses. Black contours show the posterior probability distributions for a uniform prior on $\sigsub$ with a factor of 100 uncertainty. The green and purple contours shows the results when a log Gaussian prior with 0.3 dex width is used, centered on the theoretical prediction for {\tt galacticus} and $N$-body simulations respectively.
    \label{fig:posterior_probability}}
\end{figure*}

In Figure \ref{fig:posterior_probability} we present the inferred posterior probability distributions for the hyper-parameters $\sigsub$ and $\mhm$, the normalization of the subhalo mass function and the dark matter half-mode-mass respectively. The results are marginalized over $\alpha$ and $\dlos$, the slope of the subhalo mass function and the normalization of the line-of-sight mass function, respectively, as these parameters are unconstrained by the data. 

Assuming a broad, uninformative prior on the normalization of the subhalo mass function spanning two orders of magnitude in uncertainty, we find $\log_{10}{[\mhm/M_{\odot}]}< 9.3 $ with posterior odds 10:1, and $<8.5$ at 95\% confidence.  As expected, there is a significant degeneracy between the abundance of subhalos and the warmness of dark matter. As dark matter becomes warmer, an image flux which deviates from the smooth model prediction can still be reproduced to some extent by increasing the number of more massive subhalos. The degeneracy is broken for the highest half-mode masses because these would require too many massive subhalos to reproduce the distribution of flux ratios in the full sample, particularly for lenses which are well described without additional structure beyond the macromodel.

We can make a more informative measurement by adding priors from the predicted abundance of subhalos in CDM simulations of group scale halos. To achieve this, we select values of $\sigsub$ that reproduce the bound mass function amplitude, given the tidal evolution model developed by \cite{Du++24}, plus an additional 15\% suppression due to the presence of a massive elliptical central galaxy \cite{Gannon_2025}. For a 10$^{13}$ M$_\odot$ dark matter halo at redshift z=0.5, the bound (note that \citet{Gannon_2025} referred to this as the ``evolved'' mass function, after tidal stripping) mass function amplitude is 0.05 times the infall mass function amplitude \cite{Du++25}. In the case of {\tt{galacticus}}, the value of $\Sigma_{\rm{sub}}$ that reproduces the bound mass function amplitude is $\Sigma_{\rm{sub}}=0.15 \ \rm{kpc^{-2}}$. In the case of $N$-body simulations, \citet{Gannon_2025} measure the amplitude of the bound mass function in the central 50 kpc of group scale halos in $N$-body simulations from \cite{fiacconi_cold_2016} and \cite{nadler_symphony_2023}. In this case, we find $\Sigma_{\rm{sub}}=0.1 \ \rm{kpc^{-2}}$. 
 
Both types of simulations are subject to different systematic uncertainties, including the challenges of subhalo tracking through dense environments in the case of $N$-body simulations, and the analytic treatment of tidal stripping in the case of semi-analytic simulations \citep[e.g.][]{Benson_artificial_disruption, Green++19, Griffen_caterpillar, Mansfield_symfind, Yang_galacticus_calibration}. 
Given this uncertainty, we present results based on predictions from both types of simulations, adopting broad log-normal priors with means set at the normalization value computed in the respective simulations, and a 0.2 dex uncertainty representing the relative discrepancy between the simulation types.

Adopting the prior for the projected mass in subhalos based on {\tt galacticus} simulations yields $\log_{10}{[\mhm/M_{\odot}]}< 7.8 $ with posterior odds 10:1, and $<7.3$ at 95\% confidence. Adopting a prior based on $N$-body simulations yields $\log_{10}{[\mhm/M_{\odot}]}< 7.6 $ with posterior odds 10:1, and $<7.0$ at 95\% confidence.

We can convert these limits to limits on a thermal relic particle mass $m_{\rm{therm}}$, using Equation 17 from \citet{Stucker++22} with $\beta\sim2$ and $\gamma=5$:
	\begin{equation}
		\label{eqn:mhm}
		m_{\rm{hm}} = 5 \times 10^8 \left(\frac{m_{\rm{therm}}}{3 \rm{keV}}\right)^{-10/3} \mathrm{M}_{\odot}.
	\end{equation}

\noindent This gives a lower limit on a thermal relic particle mass of 5.6 keV with the {\tt galacticus} prior, and of 6.4 keV with a prior from $N$-body simulations (10:1 odds).

\subsection{Microlensing-free flux ratios from the JWST survey}
\label{ssec:correction}
\begin{figure}
    \includegraphics[width=0.45\textwidth]{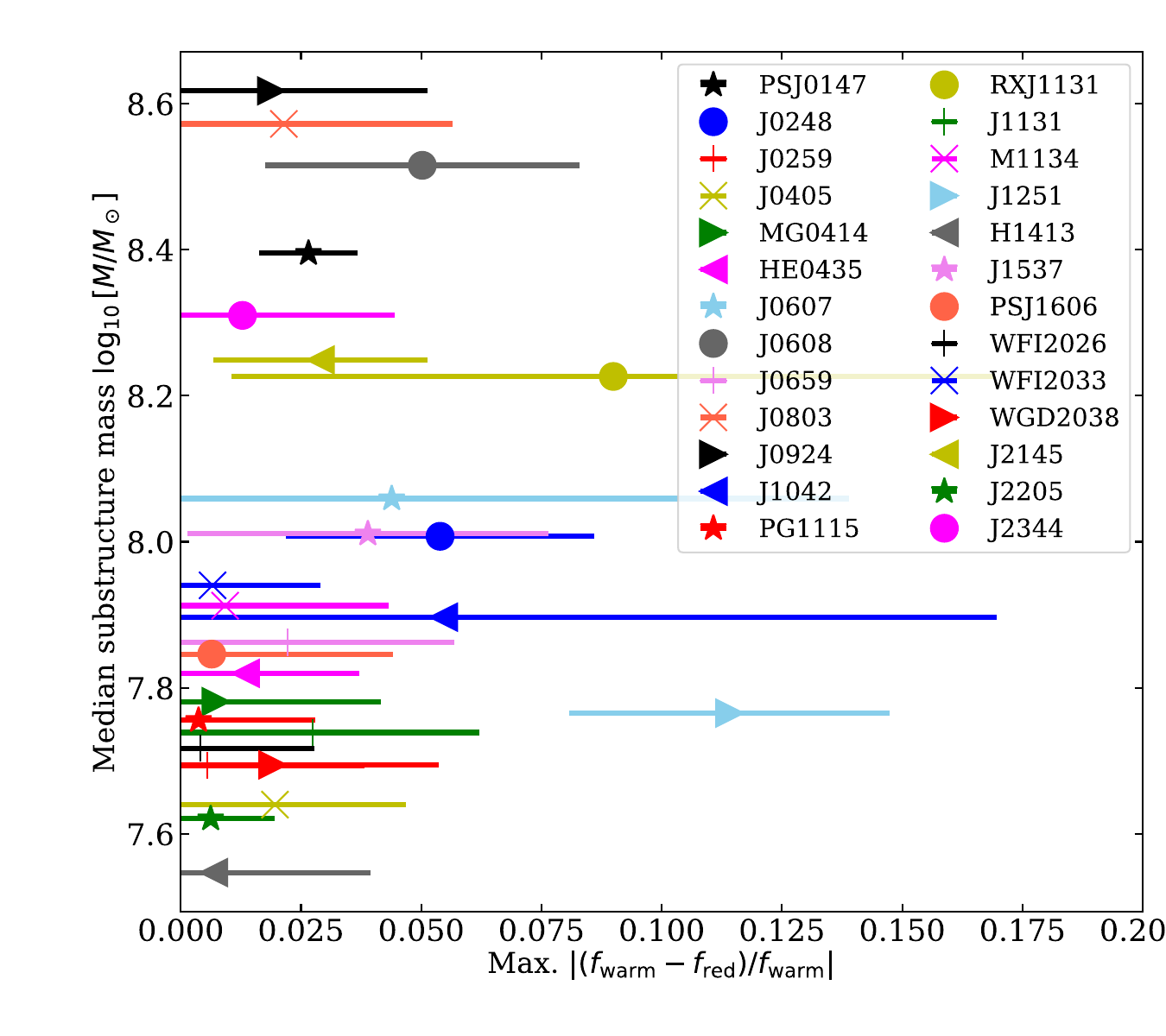}
    \caption{A comparison of the inferred mass in subhalos with the amount of correction to the flux ratios due to the SED modeling, relative to using only the reddest filter flux ratios alone (F2550W or F2100W). Systems which required more correction from SED modeling were more microlensed, so this test is intended to explore the extent to which microlensing may affect our measurement.  With the exception of J1251, the reddest filter (F2550W or F2100W) flux ratios are within two sigma of the SED-corrected flux ratios, and J1251 does not show a significant preference for additional structure. 
    \label{fig:correction_micro}}
\end{figure}
Physically, we expect no correlation between stellar microlensing and millilensing by halos. A correlation between these effects would suggest our SED fitting approach does not adequately correct for microlensing effects on the hot dust when isolating flux from the warm dust region. If this systematic were present, we would infer additional substructure in the lens model to "correct" for residual microlensing flux anomalies. 

To demonstrate that our warm dust flux ratios are free from stellar microlensing contamination, we will show that the inferred degree of millilensing is uncorrelated with the magnitude of the SED correction between the hot and warm dust regions. We define a metric for the amount of inferred millilensing based on the amount of substructure inferred for each lens system. Following the methodology summarized in Section \ref{sec:inference_framework}, we compute the likelihood function for each lens, from which we compute the posterior distribution $p\left(\Sigma_{\rm{sub}}, m_{\rm{hm}} | \datan\right)$. We draw 1000 samples from the posterior to compute the inferred total \emph{infall} mass in dark matter matter subhalos (not accounting for the effects of tidal stripping):
\begin{equation}
\langle \Sigma \rangle = \int M \frac{d^{2}N_{\rm{sub}}}{dM dA}  dM.
\end{equation}
This is simply the integral of the subhalo mass function in Equation \ref{eqn:shmf}. We fix $\alpha=-1.9$, $m_{\rm{host}}=10^{13} M_\odot$, and set $\mathcal{F}\left(M_{\rm{host}},z_{\rm{d}}\right)=1$. Higher $\langle  \Sigma \rangle$ corresponds to colder dark matter models with more subhalos, which in turn implies stronger millilensing effects perturbing the image flux ratios. 

In Figure \ref{fig:correction_micro}, we plot $\Sigma$ against the difference between the warm dust flux ratios and the flux ratios computed from the hot dust in the bluest MIRI filter. There is no trend between the magnitude of the SED correction to the warm dust and hot dust flux ratios and $\Sigma$. Therefore we do not find evidence for un-corrected microlensing influencing the dark matter inference.

\section{Discussion and conclusions}\label{sec:conclusions}
We present observations of mid-infrared emission for 31 quadruply-imaged quasars. We modeled the warm dust emission using a three-component spectral model for the continuum, hot dust torus, and more extended warm dust torus, isolating emission from the latter through a forward modeling reconstruction of the SED. The flux ratios among images from the warm dust region provide a microlensing-free dataset that we used to infer the degree of millilensing perturbations by dark matter substructure. Following the inference methodology presented in a companion paper~\citep{JWST_4_Gilman}, we interpreted these measurements within the context of a WDM model. Our main results are summarized as follows: 
\begin{itemize}
    \item We measured the image positions and multi-band fluxes and flux ratios for 31 quadruply imaged quasar lenses. We presented lens models, fluxes and flux ratios from these measurements.
    
    \item Using an updated SED fitting procedure, we infer the flux ratios of the warm dust emission for the 31 lenses in the sample. 
    
    \item We used the warm dust flux ratios in combination with narrow-line flux ratios from two additional lenses to constrain the free-streaming length of dark matter, as parameterized by the half-mode mass $m_{\rm{hm}}$. Using a prior on the normalization of the subhalo mass function from the semi-analytic simulation {\tt galacticus}, we rule out half-mode masses above $10^{7.8}$ M$_\odot$ at 10:1 odds. If we use a prior from N-Body simulations, we rule out half-mode masses above $10^{7.6}$ M$_\odot$ at 10:1 odds.
    
    \item We verify our method use of mid-IR flux ratios and our SED fitting methodology by demonstrating a lack of correlation in the inferred projected mass in substructure and the correction to the flux ratio in the reddest filter from the SED fitting.
\end{itemize}

Relative to our previous work (K24), we have made several changes which have significantly increased the complexity of the dark matter model, including allowing additional complexity in the deflector mass distribution, and significantly expanding the upper bound of the prior for the normalization of the subhalo mass function. We can approximate some of the improvement in sensitivity relative to our previous results by comparing the constraint we would have obtained using the same prior on the normalization of the subhalo mass function, $\sigsub$, as was used in K24, which corresponds to a value of $\sim 0.2$/kpc$^2$ when considered with the updated tidal stripping model used in this work. Adopting a uniform prior on $\sigsub$ with this upper limit yields $\log_{10}[\mhm]<7.3$ at 10:1 odds and $\log_{10}[\mhm]<6.8$ at 95\% confidence. This represents a stronger constraint (by a factor of 2) relative to the measurement reported in K24, indicating the relative improvement in sensitivity afforded by incorporating data from the whole sample and by the updated macromodel fitting procedure used in the dark matter inference.

These results represent a significant improvement relative to previous strong gravitational lensing measurements of dark matter properties. In a companion paper, Gilman et al. 2025 \cite{JWST_4_Gilman}, we use the measurements presented here, in conjunction with constraints on the macromodel from the lensed quasar host galaxy to obtain the strongest constraints to date on a WDM-like turnover in the halo mass function, and the most precise measurement of the normalization of the subhalo mass function around strong gravitational lenses.

The constraint presented here is $m_{\rm{therm}}>7.9$, (9.7 keV) for {\tt galacticus} or $N$-Body prior on the projected, evolved mass in substructure (95\% confidence). This is comparable to or stronger than current limits from other methods, which include $m_{\rm{therm}}>3.1$ keV (95\% confidence) 
from the Lyman-$\alpha$ forest \citep{Villasenor_2023_Lya}, 
$m_{\rm{therm}}>3.6$ keV (95\%) from stellar streams \citep{banik_stream_2018}, and $m_{\rm{therm}}>6.5$ keV (95\%) 
from Milky Way satellite galaxy counts \cite{nadler_constraints_2021}. We note that these constraints are not directly comparable between measurements, given the different prior ranges used in each analysis, particularly given the 95\% exclusion criterion.

Beyond testing for a turnover in the halo mass function, these data can be used to test a broad range of dark matter models using the general framework presented in this work and in G25.

\section*{Acknowledgments}
    We thank Alex Drlica-Wagner, Josh Frieman, and Ethan Nadler for helpful discussions. 
	
    DG acknowledges support for this work provided by the Brinson Foundation through a Brinson Prize Fellowship grant. 

    AMN, CG, MO and RK acknowledge support support from the National Science Foundation through the grant ``CAREER: An order of magnitude improvement in measurements of the physical properties of dark matter" NSF-AST-2442975.
    
    AMN, TT, XD, HP, CG, MO, RK acknowledge support from the National Science Foundation through the grant ``Collaborative Research: Measuring the physical properties of dark matter with strong gravitational lensing" NSF-AST-2205100, NSF-AST-2206315. 

    DW acknowledges support by NSF through grants NSF-AST-1906976 and NSF-AST-1836016, and from the Moore Foundation through grant 8548.

    D. Sluse acknowledges the support of the Fonds de la Recherche Scientifique-FNRS, Belgium, under grant No. 4.4503.1 and the Belgian Federal Science Policy Office (BELSPO) for the provision of financial support in the framework of the PRODEX Programme of the European Space Agency (ESA) under contract number 4000142531.

    P.M. acknowledges support from the National Science Foundation through grant NSF-AST-2407277. 

    SB acknowledges support by the Department of Physics and Astronomy, Stony Brook University

    TA acknowledges support from ANID-FONDECYT Regular Project 1240105 and the ANID BASAL project FB210003.

    KNA is partially supported by the U.S. National Science Foundation (NSF) Theoretical Physics Program Grant No.\ PHY-2210283. 

    V.N.B. acknowledges funding support from STScI grant Nos. HST-GO-17103 and HST-AR-17063. 
    
    SGD acknowledges a generous support from the Ajax Foundation.

    SFH acknowledges support through UK Research and Innovation (UKRI) under the UK government’s Horizon Europe Funding Guarantee (EP/Z533920/1, selected in the 2023 ERC Advanced Grant round) and an STFC Small Award (ST/Y001656/1).
    
    A. K. was supported by the U.S. Department of Energy (DOE) Grant No. DE-SC0009937;  by World Premier International Research Center Initiative (WPI), MEXT, Japan; and by Japan Society for the Promotion of Science (JSPS) KAKENHI Grant No. JP20H05853.
    
    V.M. acknowledges support from ANID FONDECYT Regular grant number 1231418 and Centro de Astrof\'{\i}sica de Valpara\'{\i}so CIDI 21.

    Part of this work was carried out at the Jet Propulsion Laboratory, California Institute of Technology, under a contract with NASA.

    MS acknowledges partial support from NASA grant 80NSSC22K1294. 

    K.C.W. is supported by JSPS KAKENHI Grant Numbers JP24K07089, JP24H00221.
	
	This work is based on observations made with the James Webb Space Telescope through the Cycle 1 program JWST GO-2046 (PI:Nierenberg), and the Hubble Space Telescope through HST-GO-15320, HST-GO-15652, HST-GO-17916 (PI:Treu) and HST-GO-13732 (PI:Nierenberg). Funding from NASA through this programs is gratefully acknowledged.

    Some of the data presented herein were obtained at Keck Observatory, which is a private 501(c)3 non-profit organization operated as a scientific partnership among the California Institute of Technology, the University of California, and the National Aeronautics and Space Administration. The Keck facilities we used were LRIS and OSIRIS. The Observatory was made possible by the generous financial support of the W. M. Keck Foundation.  The authors wish to recognize and acknowledge the very significant cultural role and reverence that the summit of Maunakea has always had within the Native Hawaiian community. We are most fortunate to have the opportunity to conduct observations from this mountain.

    This research is based in part on data collected at the Subaru Telescope, which is operated by the National Astronomical Observatory of Japan. We are honored and grateful for the opportunity of observing the Universe from Maunakea, which has cultural, historical, and natural significance in Hawaii.

	This work used computational and storage services provided by the University of Chicago’s Research Computing Center; Caltech's Resnick High Performance Computing Center through Carnegie Science's partnership; the Pinnacles (NSF MRI, $\#$ 2019144) and CENVAL-ARC (NSF $\#$ 2346744) computing clusters at the Cyberinfrastructure and Research Technologies (CIRT) at University of California, Merced; and the Hoffman2 Cluster which is operated by the UCLA Office of Advanced Research Computing’s Research Technology Group.  
	
	\section*{Software} This work made use of {\tt{astropy}}:\footnote{\url{http://www.astropy.org}} a community-developed core Python package and an ecosystem of tools and resources for astronomy \citep{astropy:2013, astropy:2018, astropy:2022};  {\tt{cobyqa}} \citep{rago_thesis,razh_cobyqa}; {\tt{colossus}} \citep{Diemer18};  {\tt{lenstronomy}}\footnote{\url{https://github.com/lenstronomy/lenstronomy}} \citep{birrer_lenstronomy_2018, birrer_lenstronomy_2021}; {\tt{numpy}} \citep{numpy}; {\tt{pyHalo}}\footnote{\url{https://github.com/dangilman/pyHalo}} \citep{gilman_warm_2020}; {\tt{trikde}}\footnote{\url{https://github.com/dangilman/trikde}}; {\tt{samana}}\footnote{\url{https://github.com/dangilman/samana}}; and {\tt{scipy}} \citep{scipy}. 
	
	\section*{Data availability}
	The data used in this article come from HST-GO-15320, HST-GO-15652, HST-GO-17917, HST-GO-13732 and JWST GO-2046. The raw data are publicly available online. Astrometry and and flux ratio measurements are presented by \citet{nierenberg_detection_2014, nierenberg_probing_2017, nierenberg_double_2020,Keeley_2024}. Reduced imaging data for the systems analyzed in this work are available in the open-source software {\tt{samana}}, which also provides notebooks that perform the lens modeling and scripts to reproduce the dark matter analysis. 
	
\clearpage

\bibliography{references} 

\clearpage


\appendix
\section{Deflector redshift measurements.}
\label{app:spectra}

In this section, we briefly describe the observations and redshift measurements of the lens systems J0607-2152 and J2017+6204. Spectroscopic observations for both systems were obtained with the Low Resolution Imaging Spectrometer (LRIS; \citealt{LRIS_Oke_1995}) mounted on the Keck telescopes (Program ID U175, PI: Nierenberg). The data were acquired in the long-slit mode with a 1\farcs\ wide slit using both the blue and red arms simultaneously. The blue arm was configured with a 600/4000 grism and the red arm with a 600/7500 grating, while a dichroic split the incoming light at 5600~\r{A}. This configuration provided a combined wavelength coverage from approximately 3500–9000~\r{A}, with reciprocal dispersions of around 0.6~\r{A}/pixel on the blue side and around 0.8~\r{A}/pixel on the red side. These correspond to spectral resolutions of $R \approx 1100$ on the blue side and $R \approx 1400$ on the red side. Each system was observed for a total integration time of $4 \times 2100$~s. J0607–2152 was observed in October 2023 and J2017+6204 in July 2023, both under seeing conditions of approximately 1\farcs. 

The raw data were reduced using the \textsc{LPipe} pipeline \citep{lpipe_2019} for both systems. The pipeline performed standard reduction steps including overscan subtraction, bias and dark correction, flat-fielding, cosmic-ray removal, and wavelength calibration. The resulting products included sky-subtracted 2D science frames, corresponding variance spectra, and wavelength solutions. The individual 2D science frames were then coadded using inverse-variance weighting, from which the 1D spectra were extracted. 

Due to the compact nature of the lens systems ($\sim$1\farcs) and the close proximity of bright lensed images, isolating clean 1D deflector spectra is particularly challenging in seeing-limited ground-based data. To address this, we employed the forward modeling method developed by \citet{Mozumdar_2023} to deblend the deflector spectra from quasar contamination. For J0607-2152 , we successfully deblended and extracted a high-quality deflector spectrum. For J2017+6204, where the blending was more severe, we extracted a combined (quasar + deflector) spectrum using the same method, then modeled the quasar continuum and emission lines with \textsc{PyQSOFit} \citep{PyQSOFit}. Subtracting the best-fit quasar model from the combined spectrum yielded the deflector spectrum. 

\begin{figure*}
    \includegraphics[width=0.9\textwidth]{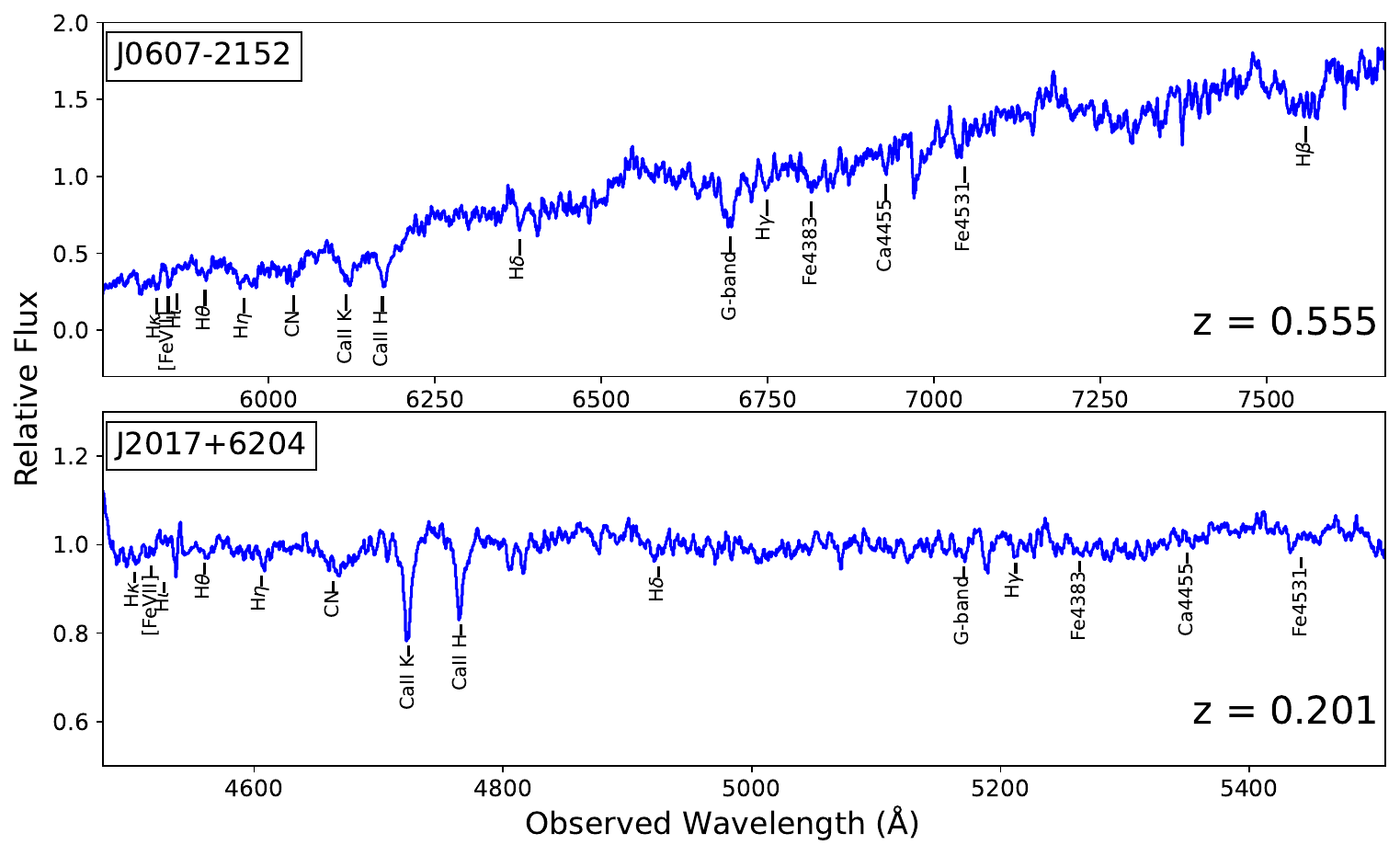}
    \caption{Deflector spectra and measured redshift for J0607-2152 (top) and J2017+6204 (bottom). The extracted and smoothed 1D spectrum of the deflector is plotted over the observed wavelength where the smoothing has been done using a boxcar filter of size around 4\r{A}. Prominent stellar absorption lines are marked if present. The measured redshift of the deflector is mentioned in the lower right corner of the plot.
    \label{fig:deflector_z}}
\end{figure*}

The resulting deflector spectra for both systems have sufficient signal-to-noise ratios, with prominent absorption features such as $\mathrm{C}$a II  $\mathrm{K\&H}$, $\mathrm{H}\delta$, and the $\mathrm{G}$-band clearly visible (Figure~\ref{fig:deflector_z}). Using these lines, we measured the deflector redshifts to be z = $0.555\pm0.001$ for J0607-2152  and z = $0.201\pm0.001$ for system J2017+6204. The quoted uncertainties correspond to the standard deviations of redshifts measured from multiple absorption features, including $\mathrm{CN}$, $\mathrm{C}$a II  $\mathrm{KH}$, $\mathrm{G}$-band, $\mathrm{H \beta}$, $\mathrm{H \delta}$, etc. We show the deflector spectra and marked the absorption lines at their measured redshifts in Fig. \ref{fig:deflector_z}.

\section{Lens models used for flux ratio measurement.}
\label{app:lensmodelparameters]}
Here we present the best fitting macromodel parameters, source and deflector light parameters and inferred PSF parameters used to measure the quasar point source fluxes for the 21 new lenses measured in this paper. Macromodel parameters and PSF parameters for the remaining 9 lenses can be found in K24. We note that these models are not used to constrain the macromodel during the dark matter inference. They are used to subtract the lensed quasar host galaxy and measure the quasar image fluxes, with the quasar image fluxes treated as independent foreground point sources. 
\input{data_tables/lens_parameters}

\input{data_tables/source_parameters}

\input{data_tables/lens_light_parameters}

\input{data_tables/psf_pars}


\section{Measured image fluxes.}
\label{app:fluxes}
Here we present the measured image fluxes for the the 21 new lenses measured in this paper. Fluxes for the remaining 9 lenses can be found in K24.

\input{data_tables/fluxes_table}


\end{document}

%% file: image_fitting_results.tex
Best-fitting model parameters for the lens mass distribution, lens light distribution, source light source light distribution, PSF, and measured quasar fluxes are given in Appendix~\ref{app:lensmodelparameters]}. Images of individual lens models are provided in the Supplementary materials.

Flux ratios in the bluest and reddest filters are given in Table~\ref{tab:data_summary}. 
Based on testing with simulated data from N24, we estimate that the absolute flux uncertainty is approximately $\sim10\%$, and that flux ratio uncertainties are $1\%$, $2\%$ or $6\%$ in the presence of no, faint, or bright lensed quasar hosts, where brightness is taken relative to the quasar point source fluxes. Figure~\ref{fig:chromaticity} shows the flux ratio variations as a function of wavelength for all lenses used in the dark matter inference.

The majority of systems were well fit by the forward modeling procedure, although we note that for several of the brightest lensed quasars, where the host galaxy is only observed in F560W, the best-fitting lens model has unphysical ellipticities. Given that our primary goal is to measure the quasar flux ratios, we chose not to restrict the lens ellipticity. Therefore, the parameters given provide a purely empirical fit to the data and are unlikely to represent the underlying mass distribution, particularly when the quasar host galaxy is only detected in F560W. These lens models are used to measure the point source fluxes in the presence of the foreground galaxy and lensed quasar host galaxy, while treating point sources as independent foreground objects. Thus these macromodels are not used to the constrain the macromodel during the forward modeling process. 

During the image fitting step we noted five systems which were unsuitable for dark matter measurements:

\emph{HE1113}: The lens galaxy was not well detected in MIRI imaging. Given its small Einstein radius of 0\farcs35, the estimated velocity dispersion is $<$150 km/s which is low enough to have a high probability of having a disk deflector. Since this cannot be ruled out with MIRI or archival HST imaging of the object, we chose to exclude this system from the dark matter analysis. 

\emph{J0457}: MIRI imaging revealed this to be a triple rather than a quad lensing system. Although in principle this could be used to constrain dark matter given the extended arc in F560W which helps to constrain the macromodel, our current forward modeling pipeline cannot accommodate three image systems so we exclude it from the current analysis. It could be incorporated in future analyses. 

\emph{B2045}: Lens models which fit the image positions predict large magnifications such that even the $\sim$10 pc warm dust region images would be extended and overlap. This is unsuitable for our current forward modeling framework which requires the lensed quasar images to be well separated to enable separate computation of the image magnifications.

\emph{J2017}: MIRI imaging revealed that the main deflector is an extended spiral galaxy. Using this system for a dark matter inference would require explicit modeling of the complex spiral morphology, making it unsuitable for dark matter measurements.

\emph{J2107}: This unusual source has a quadruply imaged broad line region \cite{2107_discovery}, but was revealed to be only double imaged in MIRI wavelengths. The bright extended arc would likely provide sufficient constraint for incorporation in future dark matter measurements, but as in the case of J0457, it cannot be modelled by our current pipeline. 

\input{data_tables/DataTables}

%% file: data_tables/DataTables.tex
\begin{table*}
    \begin{tabular}{lllllllll}
    \hline
    \hline
    Lens           & image &dRA$^a$  & dDec$^a$ &F560W Flux Ratio & Hot Ratio       & Reddest Ratio $^b$ & Warm Ratio     & [OIII] Ratio \citep{nierenberg_double_2020, nierenberg_probing_2017}\\
    \hline
     PSJ0147         &A & 0 & 0            & 0.062 $\pm$ 0.004    &0.055$^{+0.003}_{-0.005}$ & 0.0603 $\pm$ 0.0006 &0.061$^{+0.001}_{-0.001}$ \\ 
                     &B &-0.906 & 3.138  &  0.36 $\pm$ 0.02       &0.33$^{+0.01}_{-0.03}$    & 0.359 $\pm$ 0.004   &0.363$^{+0.007}_{-0.006}$ \\
                     &C &0.339 & 3.237   &      1                 &  1                       & 1                   & 1                   \\
                     &D &1.504 & 2.822  & 0.60 $\pm$ 0.04         &0.54$^{+0.03}_{-0.06}$    & 0.613 $\pm$ 0.006   &0.63$^{+0.01}_{-0.01}$ \\ 

    \hline
   
    J0248    & A    &     0    &  0 &     1      &          1                 &     1          &       1    &           \\
             &B     & -1.001 & 0.612 &1.01 $\pm$ 0.06 & 0.95$^{+0.06}_{-0.1}$ & 1.03 $\pm$ 0.02 &1.04 $^{+0.02}_{-0.02}$ \\ 
             & C    & -0.857 & 1.448 &1.25 $\pm$ 0.07 &1.22 $^{+0.07}_{-0.1}$ & 1.29 $\pm$ 0.03 &1.31 $^{+0.03}_{-0.03}$ \\ 
             &D     & 0.050 & 1.411  &0.96 $\pm$ 0.06 &0.92 $^{+0.06}_{-0.1}$ & 1.06 $\pm$ 0.02 &1.08 $^{+0.03}_{-0.02}$ \\ 
    \hline
  
    J0259  & A     &  0        &  0      &   1                 &    1                    &1              &1                  \\
               & B & -0.758 & 0.951 &1.08 $\pm$ 0.06 &1.01$^{+0.05}_{-0.08}$ & 1.02 $\pm$ 0.02 &1.02$^{+0.03}_{-0.03}$ \\ 
               & C & 0.400 & 1.265 &1.08 $\pm$ 0.06 &0.99$^{+0.06}_{-0.07}$ & 0.99 $\pm$ 0.02 &1.00$^{+0.03}_{-0.03}$ \\ 
               & D & 0.718 & 0.386 &1.62 $\pm$ 0.1 &1.46$^{+0.08}_{-0.1}$ & 1.47 $\pm$ 0.03 &1.47$^{+0.04}_{-0.04}$ \\ 
    
    \hline
    MG0414          &A    & 0     & 0     &1 &1 &1 &1\\

                  &B     & 0.132 & 0.397        &   0.90 $\pm$ 0.05   & 0.87$^{+0.01}_{-0.02}$        & 0.892 $\pm$ 0.009      & 0.89$^{+0.03}_{-0.02}$ \\
                  &C       & -0.599 & 1.944     &  0.37 $\pm$ 0.02   & 0.362$^{+0.006}_{-0.01}$       & 0.365 $\pm$ 0.004      & 0.366$^{+0.008}_{-0.009}$ \\
                  &D       &  -1.943 & 0.293    &  0.149 $\pm$ 0.009   & 0.144$^{+0.003}_{-0.003}$    & 0.144 $\pm$ 0.001      &0.143$\pm 0.003$ \\

    \hline
J0405    &A    & 1.065      & 0.325 & 1   & 1               & 1                  & 1               & 1     \\
         & B & 0 & 0&    0.68$ \pm 0.04 $&0.52$^{+0.07}_{-0.3}$ & 0.677$ \pm 0.007$ &0.69$^{+0.02}_{-0.02}$ &   $0.65\pm0.04$ \\ 
         &C    &  0.721     & 1.161& 1.14$ \pm 0.07 $&0.92$^{+0.08}_{-0.2}$ & 1.06$ \pm 0.01$ &1.06$^{+0.02}_{-0.02}$ &  $1.25\pm0.03$   \\ 
          &D    &  $-0.158$  & 1.022       1.36$ \pm 0.08 $&1.19$^{+0.09}_{-0.2}$ & 1.27$ \pm 0.01$ &1.26$^{+0.02}_{-0.03}$ & 1.26$\pm$0.02 & $1.17\pm0.04$   \\

\hline    
    HE0435          &A    & 0     & 0       &1 &1 &1 &1 &1 \\ 
                  &B & -1.475 & 0.553 &0.99$ \pm 0.06 $&0.95$^{+0.05}_{-0.08}$ & 1.01$ \pm 0.02$ &1.01$\pm 0.02$ & $1.01\pm 0.07$ \\ 
                  &C & -2.468 & -0.602 &0.85$ \pm 0.05 $&0.83$^{+0.05}_{-0.09}$ & 0.92$ \pm 0.02$ &0.93$\pm 0.02$ & $1.03\pm 0.07$ \\ 
                  &D & -0.936 & -1.615 &0.59$ \pm 0.04 $&0.57$^{+0.03}_{-0.05}$ & 0.59$ \pm 0.01$ &0.59$\pm$0.01 & 0.68$\pm$0.04 \\

    \hline
    J0457        &A &   0   &   0       & 1  & 1    & 1 & \\
                 &B    &-1.182 &-2.894 &0.84$\pm$0.02 &  0.84$^{+0.05}_{-0.03}$  &0.797$\pm$0.008 &0.78$^{+0.01}_{-0.01}$ \\
                 &C &  1.291 & 1.329 &0.79$\pm$0.02 &  0.76$^{+0.03}_{-0.03}$  &0.755$\pm$0.008 &0.76$^{+0.02}_{-0.02}$ \\
                  
    \hline
    J0607         &A    & 0            & 0    &   1      & 1                         & 1                  & 1                 \\
                  &B    & 0.140       & 1.133 & 1.18$\pm$0.07    & 1.02$^{+0.2}_{-0.5}$       & 1.42 $\pm$0.08       & 1.49$^{+0.1}_{-0.09}$    \\
                  &C    & $-0.321$    & 1.531 & 3.07$\pm0.2$     & 2.7$^{+0.6}_{-1}$         & 3.97 $\pm$0.2        & 4.17$\pm$0.3      \\
                  &D    & $-1.282$    & 0.720 & 0.93$\pm0.06$&    0.7$^{+0.2}_{-0.4}$       & 1.03 $\pm$0.06       & 1.07$^{+0.08}_{-0.07}$     \\

    \hline
        J0608    &A    & 0      & 0     &  1   & 1             & 1     & 1             &            \\
                 &B    & 0.613  & 0.603 &   0.38$ \pm 0.02 $&0.36$^{+0.01}_{-0.02}$ & 0.391$ \pm 0.004$ &0.394$^{+0.01}_{-0.008}$ \\ 
                 &C    & 1.228  &-0.273 &   0.37$ \pm 0.02 $&0.36$^{+0.01}_{-0.01}$ & 0.364$ \pm 0.004$ &0.360$^{+0.008}_{-0.008}$ \\ 
                 &D    & 0.156  &-0.394 &  0.58$ \pm 0.03 $&0.51$^{+0.03}_{-0.02}$ & 0.480$ \pm 0.005$ &0.46$^{+0.01}_{-0.02}$ \\ 

        \hline

    \hline
    J0659     &A   &  0           &  0        & 1 & 1             & 1     & 1                    \\
             &B   &  $-4.665$    &  $-0.335$ &  1.10$ \pm 0.07 $&0.94$^{+0.03}_{-0.04}$ & 0.955$ \pm 0.01$ &0.95$^{+0.02}_{-0.02}$ \\ 
             &C   &  $-0.979$    & 2.892 &    0.73$ \pm 0.04 $&0.73$^{+0.03}_{-0.02}$ & 0.701$ \pm 0.007$ &0.69$^{+0.01}_{-0.02}$ \\ 
             &D   &  0.084       & 1.903  &   3.1$ \pm 0.2 $&2.51$^{+0.08}_{-0.09}$ & 2.53$ \pm 0.03$ &2.50$^{+0.05}_{-0.06}$ \\

   \hline 
    J0803          &A    & 0     & 0     &1 &1 &1 &1\\
                  
                   &B & -0.129 & -0.345 &0.79$ \pm 0.05 $&0.80$^{+0.01}_{-0.02}$ & 0.810$ \pm 0.008$ &0.82$\pm0.02$ \\ 
                   &C & -0.976 & 0.007 &0.29$ \pm 0.02 $&0.264$\pm0.006$ & 0.260$ \pm 0.003$ &0.255$\pm0.005$ \\ 
                   & D & -0.405 & 0.810 &0.59$ \pm 0.04 $&0.563$\pm0.01$ & 0.559$ \pm 0.006$ &0.56$\pm0.01$ \\ 

    \hline
    J0924          &A    & 0     & 0     &1 &1 &1 &1\\
                   
                    & B & 0.063 & -1.806 &0.43$ \pm 0.03 $&0.43$^{+0.04}_{-0.03}$ & 0.420$ \pm 0.008$ &0.42$\pm 0.01$ \\ 
                    & C & -0.966 & -0.679 &0.37$ \pm 0.02 $&0.39$^{+0.04}_{-0.02}$ & 0.408$ \pm 0.008$ &0.42$\pm 0.01$ \\ 
                    & D & 0.529 & -0.426 &0.072$ \pm 0.004 $&0.070$\pm0.005$ & 0.133$ \pm 0.003$ &0.13$^{+0.01}_{-0.01}$ \\ 

\end{tabular}
 \caption{Results for image fitting and SED fitting for all 31 lenses. We report astrometry relative to the quasar image with dRa, dDec=0,0 and the flux ratios relative to the image with flux ratio value 1. $a$) Image position uncertainties are estimated to be 0\farcs005 for all systems, as described in Section \ref{sec:fluxes} $b$) Flux ratio of reddest filter, either F2550W or F2100W as listed in Appendix \ref{app:fluxes}.   
    \label{tab:data_summary}}
\end{table*}
\addtocounter{table}{-1}

\begin{table*}
    \begin{tabular}{lllllllll}
    Lens           & image &dRA$^a$  & dDec$^a$ &F560W Flux Ratio & Hot Ratio       & Reddest Ratio $^b$ & Warm Ratio     & [OIII] Ratio \\

\hline 

    J1042     &A   & 0      & 0        &1                      & 1                     &1                 & 1              \\
              &B   & -0.147 & -0.565    &      0.150$ \pm 0.007 $&0.17$^{+0.1}_{-0.09}$ & 0.30$ \pm 0.01$ &0.32$^{+0.03}_{-0.03}$ \\ 
              &C   & -0.817 & -0.914    &      0.077$ \pm 0.004 $&0.08$^{+0.03}_{-0.03}$ & 0.113$ \pm 0.006$ &0.115$^{+0.007}_{-0.007}$ \\ 
              &D   & -1.584 & 0.546     &      0.044$ \pm 0.002 $&0.05$^{+0.02}_{-0.02}$ & 0.072$ \pm 0.004$ &0.076$^{+0.005}_{-0.005}$ \\ 
    \hline
    HE1113        &A    & 0     & 0     &1 &1 &1 &1\\
                  &B    &-0.518 & 0.424 & 1.17$\pm$0.07 &1.1${+0.1}_{-0.3}$ & 1.20$\pm$0.01 &1.19$^{+0.01}_{-0.02}$\\  
                  &C    &-0.520 &-0.084 & 0.48$\pm$0.03 &0.42$^{+0.05}_{-0.1}$ & 0.509$\pm$0.005 & 0.510$^{+0.006}_{-0.008}$\\
                  &D    &-0.149 & 0.432 & 0.67$\pm$0.04 &0.58$^{+0.08}_{-0.2}$ & 0.707$\pm$0.007 & 0.702$^{+0.008}_{-0.01}$\\

    \hline
    
    PG1115    &A    & 0     & 0     &1 &1 &1 &1\\
              &B & 0.344 & 1.963 &1.20 $\pm$ 0.07 &1.26$^{+0.06}_{-0.09}$ & 1.33 $\pm$ 0.01 &1.35$^{+0.02}_{-0.02}$ \\ 
              &C & 1.820 & 0.382 &4.7 $\pm$ 0.3 &3.9$^{+0.4}_{-0.7}$ & 4.74 $\pm$ 0.05 &4.73$^{+0.07}_{-0.07}$ \\ 
              &D & 1.670 & -0.069 &5.0 $\pm$ 0.3 &3.6$^{+0.8}_{-1}$ & 5.36 $\pm$ 0.05 &5.30$^{+0.09}_{-0.1}$ \\

    \hline
    
        RXJ1131 &A    & 0     & 0     &1 &1 &1 &1\\
        &B & 0.028 & 1.188 &0.54 $\pm$ 0.03 &0.4$^{+0.1}_{-0.2}$ & 0.55 $\pm$ 0.01 &0.55$^{+0.03}_{-0.03}$ \\ 
        &C & -0.588 & -1.122 &0.33 $\pm$ 0.02 &0.23$^{+0.08}_{-0.1}$ & 0.447 $\pm$ 0.009 &0.41$^{+0.02}_{-0.02}$ \\ 
        &D & -3.119 & 0.877 &0.057 $\pm$ 0.003 &0.04$^{+0.01}_{-0.03}$ & 0.055 $\pm$ 0.001 &0.054$^{+0.003}_{-0.003}$ \\

    \hline
    GRAL1131     &A    & 0     & 0     &1 &1 &1 &1\\
             
             &B & -0.940 & 1.086 &1.62$ \pm 0.1 $&1.1$^{+0.4}_{-0.6}$ & 1.56$ \pm 0.03$ &1.52$^{+0.03}_{-0.04}$ \\ 
             &C & 0.361 & 1.520 &3.8$ \pm 0.2 $ &3.1$^{+0.6}_{-2}$ & 4.06$ \pm 0.08$ &4.00$^{+0.07}_{-0.08}$ \\ 
             &D & 0.689 & 1.185 &3.7$ \pm 0.2 $ &3.6$^{+0.2}_{-0.5}$ & 4.05 $ \pm 0.08$ &4.00 $\pm0.07$ \\ 

    \hline    
   
    2M1134          &A    & 0     & 0     &1 &1 &1 &1\\
                
                 & B & 0.700 & 2.141 &0.27$ \pm 0.02 $&0.24$^{+0.02}_{-0.02}$ & 0.246$ \pm 0.002$ &0.244$^{+0.004}_{-0.004}$ \\ 
                 & C & 2.678 & 2.531 &1.09$ \pm 0.07 $&1.04$^{+0.06}_{-0.06}$ & 1.05$ \pm 0.01$ &1.05$^{+0.02}_{-0.02}$ \\ 
                 & D & 1.948 & 0.774 &0.71$ \pm 0.04 $&0.53$^{+0.03}_{-0.04}$ & 0.548$ \pm 0.005$ &0.544$^{+0.01}_{-0.01}$ \\ 

        \hline     
    J1251         &A    & 0     & 0     &0.25 &0.21$^{+0.09}_{-0.14}$ &0.217 &0.193$^{0.012}_{-0.018}$\\
                  &B    & 1.713 & 0.016 &0.46 &0.61$^{+0.22}_{-0.10}$  &0.458 &0.455$^{+0.016}_{-0.014}$\\  
                  &C    & 1.782 &-0.572 &1 &1 &1 &1 \\
                  &D    & 1.423 &-0.947 &0.67 &0.59$^{+0.24}_{-0.36}$ &0.575 &0.541$^{+0.028}_{-0.044}$ \\

    \hline    
    H1413          &A    & 0     & 0      &1.18  &1.14$\pm$0.04  & 1.15 &1.15$\pm$0.02\\
                  
                  & B & 0.746 & 0.168 &0.86$ \pm 0.05 $&1.14$^{+0.03}_{-0.05}$ & 0.872$ \pm 0.009$ &1.15$^{+0.02}_{-0.02}$ \\ 
                  & C & -0.489 & 0.712 &0.78$ \pm 0.05 $&0.95$^{+0.03}_{-0.04}$ & 0.845$ \pm 0.008$ &0.98$^{+0.01}_{-0.01}$ \\ 
                  & D & 0.357 & 1.041 &0.43$ \pm 0.03 $&0.48$^{+0.01}_{-0.02}$ & 0.426$ \pm 0.004$ &0.484$^{+0.007}_{-0.007}$ \\

      \hline
            
           J1537           &A   & 0     & 0      &1      &    1                   & 1       & 1             &            \\
                           &B   &-1.993 &-0.329  &    0.81$ \pm 0.05 $&0.5$^{+0.2}_{-0.3}$ & 0.73$ \pm 0.01$ &0.73$^{+0.02}_{-0.02}$ \\ 
                           &C   &-2.848 & 1.644  & 1.07$ \pm 0.06 $&1.5$^{+0.6}_{-0.3}$ & 0.99$ \pm 0.02$ &0.96$^{+0.03}_{-0.03}$ \\ 
                           &D   &-0.750 & 1.763  &0.75$ \pm 0.05 $&0.7$^{+0.2}_{-0.3}$ & 0.73$ \pm 0.01$ &0.73$^{+0.02}_{-0.02}$ \\

     \hline
     J1606             &A   & 0      & 0      &1    & 1              & 1     & 1           &  1.00$\pm 0.03$ \\
                       &B   & -1.621 & -0.592&   1.01$ \pm 0.06 $&0.97$^{+0.05}_{-0.09}$ & 1.01$ \pm 0.02$ &1.01$^{+0.02}_{-0.02}$ & 1.00$\pm 0.03$ \\
                       &C   & -0.792 & -0.905 &  0.60$ \pm 0.04 $&0.57$^{+0.03}_{-0.06}$ & 0.59$ \pm 0.01$ &0.58$^{+0.01}_{-0.01}$ &  0.60$\pm0.02$ \\ 
                       &D   & -1.129 & 0.152  & 0.73$ \pm 0.04 $&0.74$^{+0.04}_{-0.05}$ & 0.75$ \pm 0.02$ &0.75$^{+0.01}_{-0.01}$ &  0.78$\pm 0.02$ \\
\hline

    J2017          &A    & 0     & 0     &1 &1 &1 &1\\
                  &B    & 0.475 & 0.436 & 1.92 &2.2$\pm$0.1    & 2.08 &2.10$\pm$0.05\\  
                  &C    & 0.872 & 0.135 & 1.47 &1.56$\pm$0.08  & 1.52 &1.55$\pm$0.04 \\
                  &D    & 0.585 &-0.712 & 0.72 &1.08$\pm$0.06  & 1.07 &1.14$\pm$0.03 \\ 
   
    \hline

    J2026   &A1  &  0     & 0     &1                 & 1             & 1       & 1                     &  1.00$\pm$0.02    \\
            &A2  &  0.252 & 0.219 & 0.74$ \pm 0.04 $&0.73$^{+0.03}_{-0.04}$ & 0.772$ \pm 0.008$        &  0.77$^{+0.01}_{-0.01}$     &  0.75$\pm$0.02 \\  
            & B  & -0.164 & 1.431 & 0.31$ \pm 0.02 $&0.32$^{+0.01}_{-0.01}$ & 0.303$ \pm 0.003$        &  0.304$^{+0.005}_{-0.005}$  & 0.31$\pm$0.02 \\  
            &C   & -0.733 & 0.386 &0.28$\pm 0.02$   &0.28$ \pm 0.02 $        &0.274$^{+0.009}_{-0.01}$ &  0.281$^{+0.004}_{-0.004}$ &  0.28$\pm$0.02 \\ 
    \hline
    
    \end{tabular}
    \caption[]{Continued.}   
\end{table*}
\addtocounter{table}{-1}

    \begin{table*}
    \begin{tabular}{lllllllll}
    \hline
    \hline
    Lens           & image &dRA$^a$  & dDec$^a$ &F560W Flux Ratio & Hot Ratio       & Reddest Ratio $^b$ & Warm Ratio     & [OIII] Ratio \\
    \hline
    WFI2033          &A1    & 0     & 0     &1 &1 &1 &1 &1\\
                 
                  & A2 & 0.713 & 0.117 &0.70$ \pm 0.04 $&0.60$^{+0.06}_{-0.1}$ & 0.718$ \pm 0.007$  &0.72$^{+0.01}_{-0.01}$ & 0.64 $\pm$0.03 \\ 
                  & B & 2.198 & -1.259 &0.52$ \pm 0.03 $&0.45$^{+0.04}_{-0.09}$ & 0.526$ \pm 0.005$ &0.527$^{+0.009}_{-0.008}$ & 0.50$\pm$0.02 \\ 
                  & C & 0.084 & -1.540 &0.64$ \pm 0.04 $&0.55$^{+0.05}_{-0.1}$ & 0.609$ \pm 0.006$  &0.61$^{+0.01}_{-0.01}$ & 0.53 $\pm 0.02$\\ 
    \hline

    J2038     &A   & 0      & 0      & 1  & 1                    & 1      & 1               &  1$\pm$0.01  \\
              &B    & 2.307  & -1.707 & 1.16$ \pm 0.07 $&0.9$^{+0.2}_{-0.5}$ & 1.23$ \pm 0.01$ &1.22$^{+0.02}_{-0.04}$ &  1.16$\pm$0.02 \\
              &C   & 0.796  & -1.678 &0.91$ \pm 0.05 $&0.8$^{+0.2}_{-0.4}$ & 0.956$ \pm 0.01$ &0.95$^{+0.02}_{-0.03}$ &  0.92$\pm$0.02 \\
                &D   & 2.178  &  0.384 & 0.42$ \pm 0.03 $&0.35$^{+0.08}_{-0.2}$ & 0.438$ \pm 0.004$ &0.435$^{+0.008}_{-0.01}$  &  0.46$\pm$0.01 \\

    \hline
    
    B2045         &A &0      &0      &1     &1 &1 &1\\
                  &B &-0.131 &-0.242 &0.57$\pm$0.03  &0.63$^{+0.08}_{-0.03}$ &0.532$\pm$0.005 &0.50$\pm$0.02  \\
                  &C &-0.288 &-0.791 &0.69$\pm$0.04  &0.61$^{+0.05}_{-0.16}$ &0.710$\pm0.007$ &0.73$\pm$0.02 \\
                  &D$^{c}$ & 2.900 &-1.549 &0.017$\pm$0.002 &-  &-  &-\\
 
   \hline
        
                 J2107  &A &0      &0      &1    & 1 &1    &1 \\
                         &B &-3.680 &-0.835 & 0.66 $\pm$ 0.04 &0.72$^{+0.1}_{-0.08}$ & 1.33 $\pm$ 0.03 &1.4$^{+0.2}_{-0.2}$ \\ 

    \hline

     J2145         &A     &0     &0         &   0.230 $\pm$ 0.002       &0.228$^{+0.004}_{-0.01}$   & 0.242 $\pm$ 0.002 &0.248$^{+0.002}_{-0.002}$ \\ 
                   &B    & 0.465 & 1.672    &   0.138 $\pm$ 0.001       &0.138$^{+0.002}_{-0.003}$  & 0.135 $\pm$ 0.001 &0.140$^{+0.001}_{-0.001}$ \\ 
                   &C    & 1.843 & 0.927    &   0.673 $\pm$ 0.007       &0.65$^{+0.02}_{-0.06}$     & 0.755 $\pm$ 0.008 &0.759$^{+0.008}_{-0.008}$ \\ 
                   &D    & 1.530 & 0.358    & 1                         &1                          &1                  &1                          \\
    \hline
    J2205          &A    & 0     & 0     &1 &1 &1 &1\\
            
                  & B & -1.199 & -0.725 &1.52$ \pm 0.03 $&1.48$^{+0.05}_{-0.08}$ & 1.54$ \pm 0.02$ &1.55$^{+0.01}_{-0.01}$ \\ 
                  & C & -1.642 & -0.121 &2.81$ \pm 0.06 $&2.6$^{+0.2}_{-0.3}$    & 2.71$ \pm 0.03$ &2.70$^{+0.03}_{-0.03}$ \\ 
                  & D & -1.349 & 0.439 &1.66$ \pm 0.03 $&1.59$^{+0.06}_{-0.1}$   & 1.67$ \pm 0.02$ &1.67$^{+0.02}_{-0.01}$ \\

    \hline
    J2344           &A & 0 & 0 &1 &1 &1 &1\\
                   
                   & B & 0.585 & 0.347 &1.36$ \pm 0.08 $&0.8$^{+0.4}_{-0.4}$ & 1.18$ \pm 0.02$ &1.17$^{+0.03}_{-0.03}$ \\ 
                   & C & 0.876 & -0.323 &1.04$ \pm 0.06 $&0.6$^{+0.3}_{-0.3}$ & 0.91$ \pm 0.02$ &0.90$^{+0.02}_{-0.02}$ \\ 
                   & D & 0.241 & -0.661 &1.48$ \pm 0.09 $&1.0$^{+0.4}_{-0.5}$ & 1.50$ \pm 0.03$ &1.49$^{+0.03}_{-0.03}$ \\

    \hline
    \end{tabular}
    \caption[]{Continued. $c$) Image D of J2045 not detected in all filters.}   
\end{table*}

%% file: sed_fitting_method.tex
This section describes the procedure we use to fit the near to mid infrared spectral energy distribution (SED) of the lensed quasar images in order to measure the flux emitted from the warm dust region, that can be used for dark matter measurements, because it is not affected by microlensing. Stars in the lensing galaxy have characteristic lensing size scales of order $\sim \mu$as, therefore, it is important that the source light used for dark matter measurements be much larger than this so that it is not differentially magnified by stars. 

At rest-frame near-IR ($\sim1-3\mu$m), the quasar SED is primarily composed of light emitted from the quasar accretion disk, which has a size of $\sim\mu$as, as well as from the dust surrounding the quasar. The hottest dust, with a typical color temperature of $\sim 1500$ K \cite[e.g.][]{Poindexter_2007, Mor_sed_2012}, is associated with the sublimation zone. This hot dust component dominates the SED flux at wavelengths of $\sim 3 \mu$m. 
The dust sublimation zone, corresponding to the innermost region from the central engine in which dust is not sublimated, depends somewhat on dust composition and strongly on the luminosity of the quasar.  
This region can be as small as 0.05--0.2 pc for low quasar luminosities of $10^{44}-10^{45}$~erg/s \citep{suganuma_reverberation_2004, suganuma_reverberation_2006, mor_hot-dust_2011,gravity_collaboration_image_2020}. 
This small size makes the hot dust susceptible to microlensing for some lenses \citep{sluse_mid-infrared_2013}. 

Dust emitting at rest-frame wavelengths of $\sim$8--12$\mu$m is typically referred to as the `warm' dust with temperatures of $\sim$ hundreds of Kelvin, and is typically seen to occupy regions between $\sim0.5-10$ pc with little to no dependence on quasar luminosity \citep{burtscher_diversity_2013, leftley_parsec-scale_2019, honig_redefining_2019}, making it $\sim$mas in extent, and therefore not susceptible to microlensing. 

We estimated the contribution of the `hot' dust to the reddest observed MIRI band using the three component SED fitting procedure described in N24. The emission from the quasar was modeled as a power-law $\nu F_\nu \propto \nu^{\beta_q}$ where the slope $\beta_q$ was allowed to vary between 0 and -0.8. This extreme range around the typical observed value of -0.3 was selected to account for distortions in the continuum due to microlensing. The hot and warm dust were fit as two blackbodies with variable temperatures. The relative amplitudes of the continuum, hot dust, and warm dust were allowed to vary between the lensed images to account for differential microlensing of the different size regions. Additional details of the inference methodology are provided by N24 and K24. An example of the three component fit to the quasar SEDs for the lens J0924 is shown in Figure \ref{fig:sed_0924_main}.

We made a small improvement in the fitting procedure relative to K24; previously we had adopted a broad prior for the hot dust temperature, allowing it to vary between 900-1600 K based on the results of \citet{Hernan-Caballero-mid-ir-2016}. From the larger sample in this work, we noted several systems with un-physical solutions in which the warm dust flux went to zero. The goal of including a hot dust component in the SED fitting model was to account for dust that may be microlensed. Therefore we updated the prior to only allow the hot dust blackbody to have a lower limit of 1200 K (rather than 900 K). We applied this new prior and re-fit all 31 lenses in the sample.  Therefore, the warm dust flux ratios inferred in this paper supersede values proposed in K24. The primary effect of this change was to reduce the measurement uncertainties, as the potential impact of variations in the hot dust component on longer wavelengths was reduced by restricting this component to have a higher blackbody temperature.

For each lens, we fit the multi-component SED for the four lensed quasar images, keeping the temperatures of the two blackbodies fixed while allowing the amplitudes of the blackbodies and the continuum emission as well as the continuum slope to vary, to account for differential microlensing and source variability.  The SED-predicted fluxes for the normalization image (set to 1 in Table \ref{tab:data_summary}, and flux ratios for the remaining images were compared to the data compute a likelihood for each realization. Uncertainties in the fluxes and flux ratios are dominated by uncertainties in the PSF and quasar host galaxy model, and approximated as being Gaussian, based on N24. 
The posterior probability distribution was sampled with {\tt emcee} \citep{foreman-mackey_emcee_2013}. The mean parameter values and covariance for the flux ratios were estimated from one thousand random draws from the samples of the Markov Chain Monte Carlo.

%% file: sed_fitting_results.tex
\begin{figure*}
    \includegraphics[width=\textwidth]{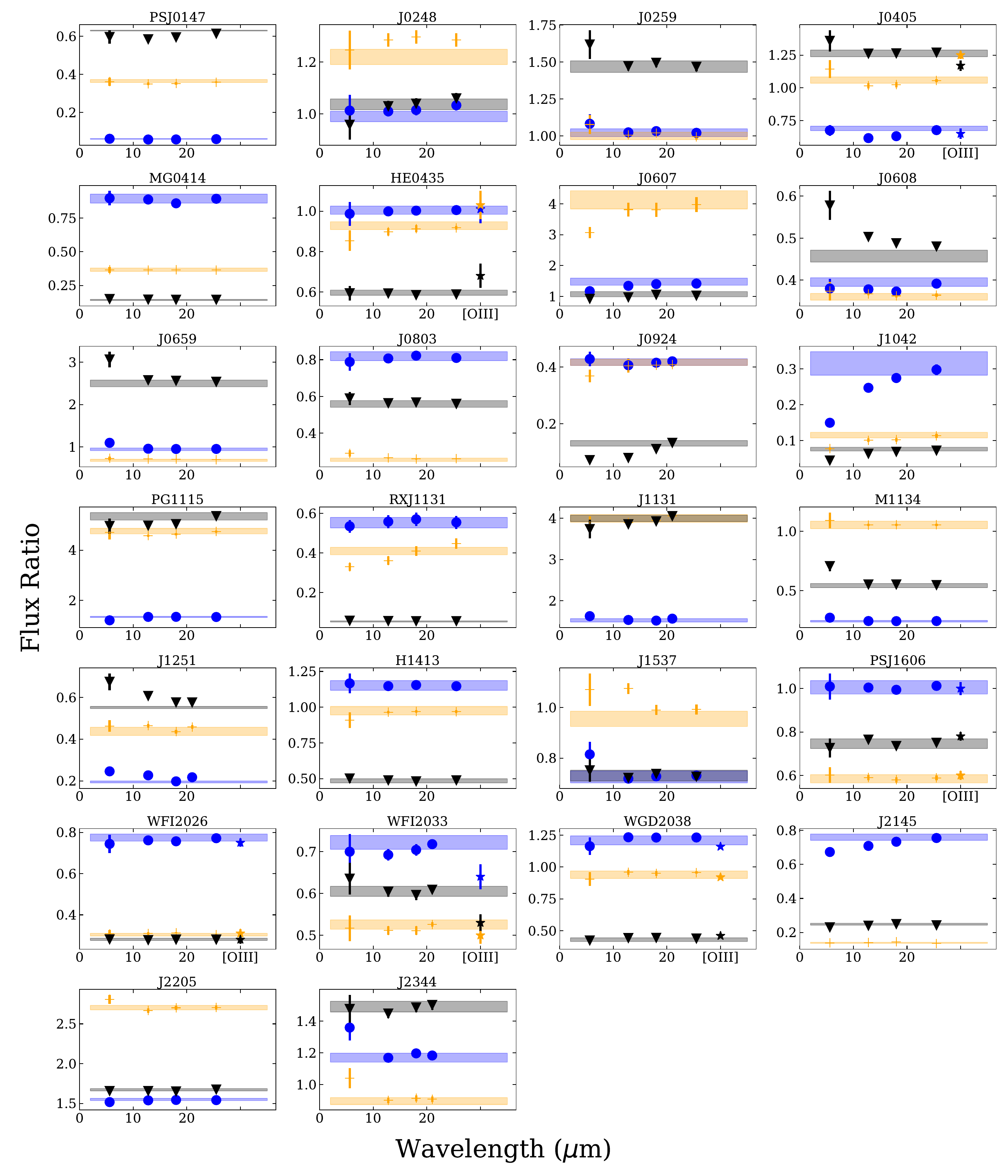}
    \caption{Comparison of multi-band SED fitting result (filled bands) with the flux ratios in each filter for the 26 lenses used in the dark matter inference. Blue circles, orange crosses and black triangles correspond to B/A, C/A and D/A respectively for all lenses, with the exception of: PSJ0147 and J1251 (A/C, B/C, D/C),  H1413 (A/B, C/B, D/B) and WFI2033 (A2/A1, B/A1, C/A1). For all lenses, with the exception of J1251, the flux ratio in the reddest filter is less than 2 $\sigma$ from the inferred warm dust flux ratio, consistent with minimal microlensing contamination. For systems with narrow-line flux ratio measurements, we show the narrow-line measurements as stars at an arbitrary wavelength for comparison. \label{fig:chromaticity}}
\end{figure*}

Table~\ref{tab:data_summary} provides the inferred warm and hot dust flux ratios and uncertainties resulting from the SED fitting, while Figure~\ref{fig:chromaticity} shows how the flux ratios in each MIRI band vary with wavelength, compared to the inferred warm dust flux ratio, and to the narrow-line flux ratios, if available from the literature \cite{nierenberg_detection_2014, nierenberg_probing_2017, nierenberg_double_2020}. 

Figure~\ref{fig:chromaticity} shows that several systems have significant microlensing in the bluest filter, F560W, with the flux ratio in at least one image more than two sigma away from the inferred warm dust flux ratio. Specific examples are  J0607, J0608, J0924, J1251 and J2344. 

J0924 has long been known as a case of extreme microlensing. Given its fold configuration, it is expected that the two merging images (A and D) should have similar brightnesses, and yet in optical bands image D is observed to be much fainter than A.  There is significant variation in the relative flux of image A and D with wavelength, ranging from a flux ratio of D/A of 0.04 in HST ACS WFC F555W ($\sim$0.536$\mu$m) to about 0.08 in HST NICMOS F160W ($\sim$1.6$\mu$m), with approximately 20\% measurement uncertainties \citep{Morgan_0924_2006}. Interestingly, this flux ratio has been approximately constant over the past two decades, contrary to expectations if the anomalous flux ratio were caused entirely by microlensing \cite{Morgan_0924_2006}. We measure a flux ratio of D/A of 0.073$\pm$0.004 in MIRI F560W and a warm dust flux ratio of 0.13$\pm$0.01. This is still highly anomalous relative to the smooth model prediction of $\sim$1 for this flux ratio. 

Interestingly, the warm dust measurement in J0924 is very different from a recent measurement of the flux ratios for this system in the radio: using the CO(5-4) emission line, \citet{badole_vla_2020} finds D/A$\sim$1.07. \citet{badole_vla_2020} also estimates that the intrinsic size of the emitting region is between 850-2500 pc, which is 1 to 2 orders of magnitude larger than the warm dust region. Therefore, it is possible that the discrepancy is due to a perturbation on the size scale comparable to that of the warm dust region, but which is too small to also perturb the CO emitting region. If so, this would place a strong constraint on the size-scale of the perturbation.

\begin{figure*}
    \includegraphics[width=0.45\textwidth]{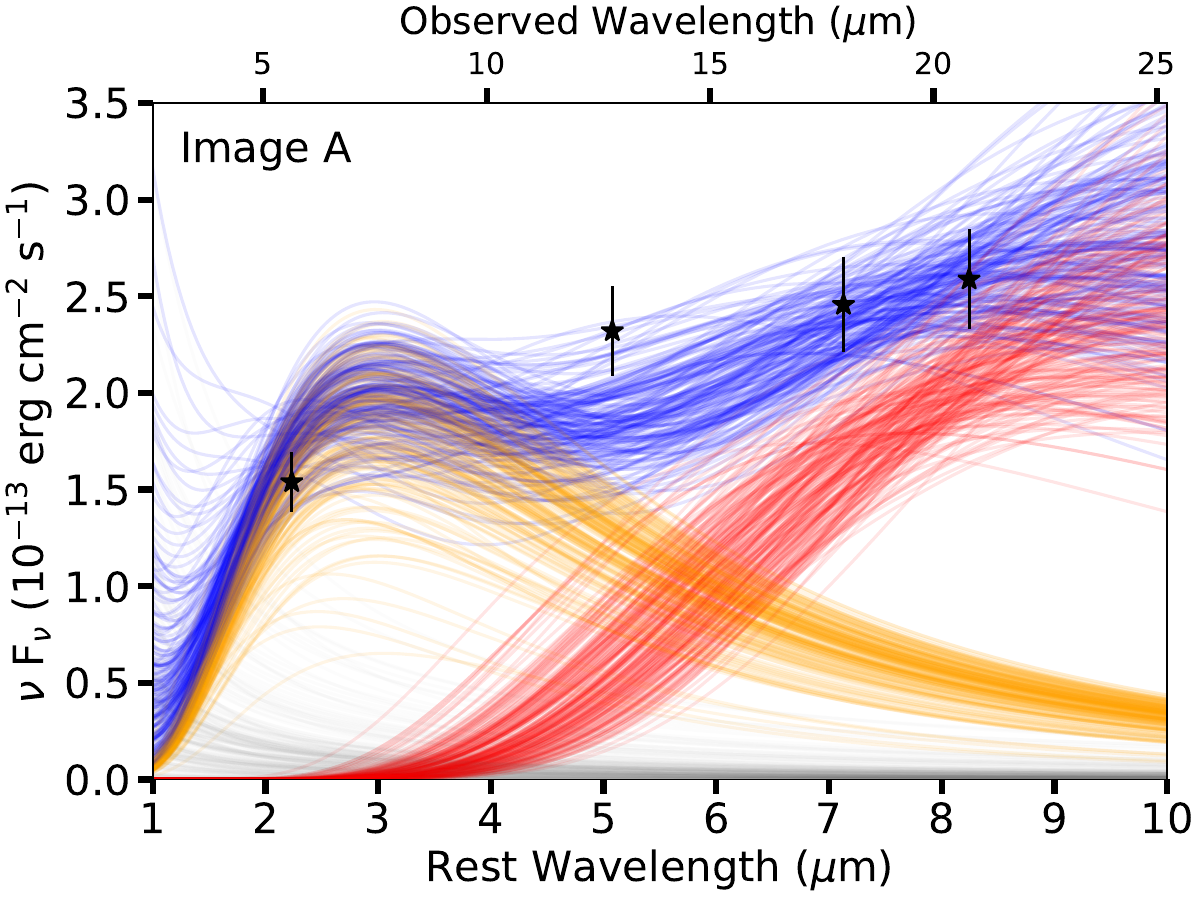}\includegraphics[width=0.45\textwidth]{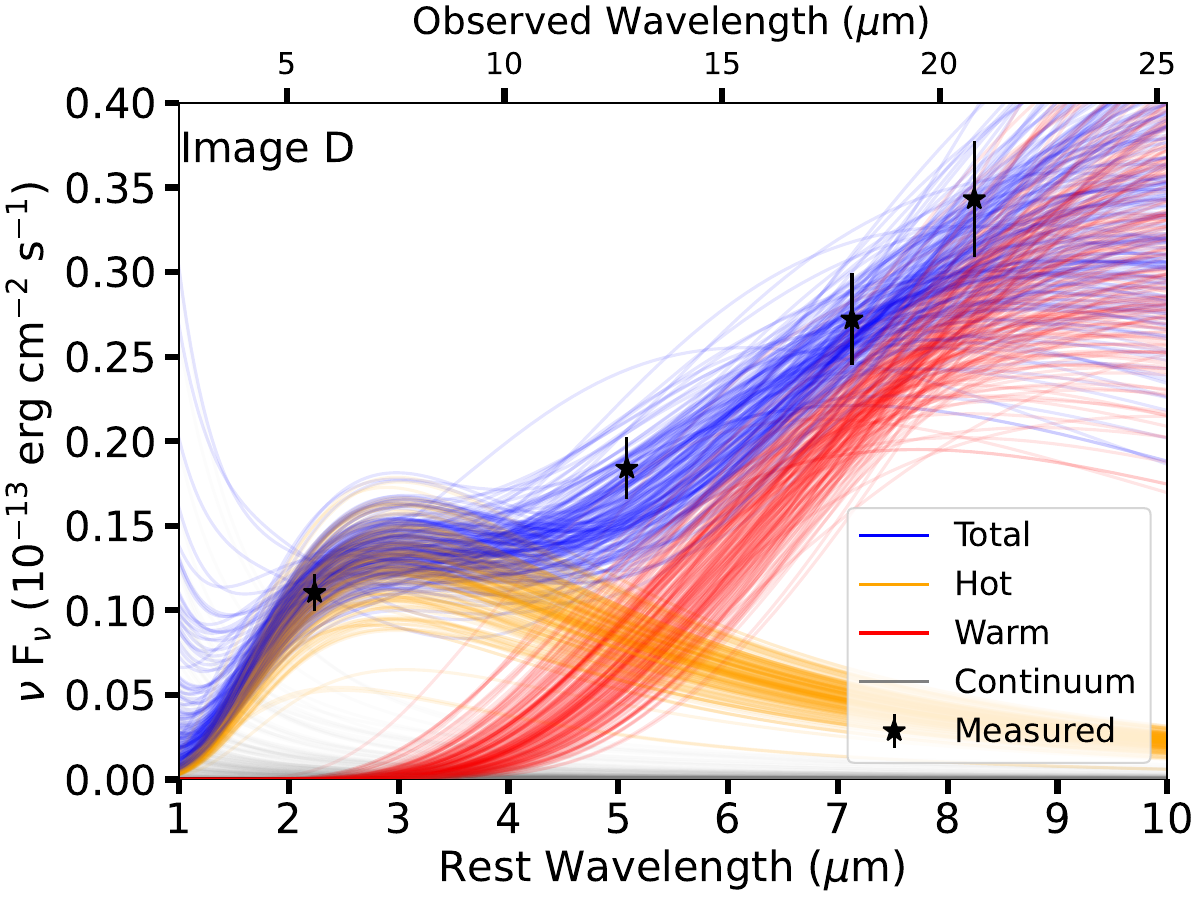}
    \caption{The results of the spectral energy distribution fitting for J0924 for two of the four lensed images. The extreme difference in the spectral energy distribution model between the lensed images reveals strong microlensing in this system. The smooth macromodel expectation based on the image positions is that image A and D should have the same flux. J0924 shows a significant deviation from this expectation.
    \label{fig:sed_0924_main}}
\end{figure*}

We note that the goodness of fit of the two blackbody plus continuum model is lower for J0924 than for other lenses. In particular, the F1280W filter is less well matched than for other systems. Figure \ref{fig:sed_0924_main} shows a comparison of the SED fit to images A and D for this lens.

Although this filter is matched within the measurement uncertainties of the model, it may be that the model is insufficiently flexible to fully account for the microlensing in image D. We note, however, that the measured F2100W flux can be viewed as an upper limit on the possible total flux of this image, and therefore the system is strongly perturbed relative to the smooth model expectation.

For four of the six lenses, the narrow-line ([OIII]) flux ratios are consistent with the inferred warm dust flux ratios within the 68\% confidence intervals. J0405 and WFI2033 show somewhat larger differences although the difference in the flux ratios for WFI2033 is less than $2\sigma$. As noted by N24, the discrepancy between flux ratios in J0405 may be due to differential milli-lensing by a small subhalo that affects the smaller warm dust emission more than the more extended narrow-line emission. Alternatively, this may indicate an under-estimation of measurement uncertainties for one of the methods. We leave further investigation of the difference in flux ratios for this system to a future work.

In general, the SED fitting provides a small correction to the flux ratio in the reddest filter, with only one system, J1251 showing a larger than $\sim$ 2$\sigma$ difference between the warm dust and reddest filter flux ratios.


%% file: data_tables/lens_parameters.tex
\begin{table*}
    \centering
    \begin{tabular}{lllllllll|lll}
    \hline
    \hline
    Lens          & $\theta_{\rm{E}}$ & $\gamma_p$ & $q$ &  $\phi$ (degree) & $\gamma_{\rm{ext}}$ & $\phi_\gamma$ (degree) & dRA& dDec&    $\theta_{\rm{G2}}$ & dRA$_{\rm{G2}}$ &dDec$_{\rm{G2}}$  \\
    \hline
    
    \hline
    J0248 & 0.76 & 2.13 & 0.64 & 330 & 0.174 & 344 & -0.364 & 0.820 \\
    \hline
    J0259  &0.73 & 2.04 & 0.66 & 316 & 0.022 & 19 & 0.033 & 0.668     \\
    \hline
    MG0414 & 1.04 & 2.26 & 0.60 & 274 & 0.148 & 353 & -1.182 & 0.637 \\
    \hline
    HE0435       & 1.20 & 2.07 & 0.56 & 373 & 0.011 & 55 & -1.182 & -0.575 &   &  & \\
    \hline
    J0457 & 1.82 & 2.00 & 0.54 & 348 & 0.189 & 221 & 2.160 & -2.213\\
    \hline
    J0803  &0.57 & 1.88 & 0.96 & 327 & 0.120 & 31 & -0.579 & 0.113 \\
    \hline
    J0924          &  0.88 & 2.12 & 0.96 & 319 & 0.054 & 53 & -0.188 & -0.866   \\
    \hline 
    HE1113$^a$   & 0.12 & 2.00 & 1.00 & 199 & 0.541 & 131 & -0.178 & 0.122\\
    \hline
    PG1115 & 1.15 & 1.90 & 0.80 & 153 & 0.091 & 278 & 0.696 & 0.570 \\
    \hline
    RXJ1131 &  1.81 & 1.94 & 0.89 & 289 & 0.116 & 232 & -1.991 & 0.562\\
    \hline
    GRAL1131&  0.87 & 2.04 & 0.62 & 190 & 0.056 & 178 & -0.064 & 0.687\\
    \hline
    2M1134  & 1.24 & 2.29 & 0.52 & 255 & 0.382 & 173 & 1.200 & 1.537 \\
    \hline
    J1251     &  0.84 & 1.78 & 0.78 & 278 & 0.041 & 208 & 1.087 & -0.342\\
    \hline
    H1413 & 0.66 & 1.70 & 0.38 & 227 & 0.168 & 236 & 0.199 & 0.590 \\
    \hline
    J2017 & 0.59 & 2.03 & 0.32 & 313 & 0.103 & 302 & 0.379 & -0.056    \\
    \hline    
    WFI2033 &  1.10 & 1.71 & 0.79 & 292 & 0.092 & 77 & 0.710 & -0.959& 1.10 & 1.095 & 1.127\\
    \hline  
    B2045 & 1.91 & 2.00 & 0.17 & 10 & 0.312 & 9 & 2.902 & -1.553\\
    \hline
    J2107 &2.11 & 2.00 & 0.74 & 361 & 0.062 & 325 & -1.987 & -1.475\\
    \hline
    J2205        & 0.80 & 1.72 & 0.58 & 20 & 0.060 & 23 & -0.951 & -0.112 \\
  \hline
     J2344  & 0.51 & 1.70 & 0.65 & 63 & 0.063 & 49 & 0.460 & -0.165\\
  \hline
 
    \end{tabular}
    \caption{Best fitting lens model parameters, with the image positions not used in the lens model. Position angles are given in degrees East of North, all other angles are in units of arcseconds. Lens centers given relative to coordinate system in Table \ref{tab:data_summary}.
    Lens J2145 did not have detected extended arcs in any filter and thus no lens model was applied. (a) This lens model incorrectly produces a fourth (counter) image which is not observed. It was, however, suitable for our purpose of removing the flux of the lensed quasar host galaxy to enable accurate point source measurements. }
    
    \label{tab:macromodel}
\end{table*}

%% file: data_tables/source_parameters.tex
\begin{table*}
    \centering
    \begin{tabular}{llll|ll|lll}
    \hline
    \hline
    Lens    &Filter & r$_{\rm{s}}$ (arcsec)& n$_{\rm{s}}$ & n$_{\rm{max}}$ &  $\beta$ (arcsec) & dRa & dDec\\
    \hline
    PSJ0147          &F560W  & 0.289 & - & -  & 0.50 & 0.807 & 0.069 \\
    \hline
    J0248 & F560W    & 0.252 & 2.34 &   3   &  1.2    & $-$0.777 & 1.120 \\
          & F1280W   & 1.095 & 2.76 &   -    &    -    &       &        & \\
          & F1800W    & 0.394 & 0.50 &  -     &   -     &       &        &  \\
          & F2550W   & 0.628 & 1.54 &   -    &    -    &       &        & \\
    
    \hline
    J0259   & F560W  & 0.130 & 0.87 & -     &  -   & $-$0.070 & 0.645  \\ 
            & F1280W & 0.096 & 3.24 &  -    &  -   &        &        \\
            & F1800W & 0.043 & 2.83 &  -    &  -   &        &        \\
            & F2550W & 0.044 & 0.68 &  -    &  -   &        &        \\
    
    \hline
    
    MG0414 &F560W  &0.331 & 4.00 &    -           &      -         &   $-$1.267 & 1.139 \\
    \hline
   
    HE0435   & F560W & 0.256 & 1.63  &   3    &    1.3      &      $-$1.169 & $-$0.555    \\
             & F1280W & 0.263 & 4.00 &  -   &     -        &             &            \\
             & F1800W & 0.474 & 2.16 &   -   &    -         &             &            \\
             & F2550W &0.248 & 0.64  &  -    &     -        &             &            \\

    \hline
    J0457    &F560W   &0.36 & 1.0 & -&- & 2.724 & $-$0.997 \\
    \hline
    J0803     & F560W & 0.102 & 3.51 & -   & -     &    $-$0.899 & 0.260 \\     
    
    \hline
    
    J0924   & F560W   &    0.189       & 3.32 & 3  & 1.1  &  $-$0.373 & $-$0.876 \\
            & F1280W  &    0.227       & 0.50 & -  &  -     &         &       \\
            & F1800W  &    0.244       & 0.56 & -  &   -    &         &       \\
            & F2100W  &    0.242       & 0.50 & -  &   -    &         &       \\
   
    \hline
    HE1113 &F560W &0.20 & 1.0 & - & - & $-$1.857 & 0.796 \\
    \hline
    PG1115 &F560W &0.063 & 4.00 &   &  &0.974 & 1.197  \\
    \hline

RXJ1131 source 1 & F560W & 0.635 & 4.00 & 11 & 0.2 & 0.722 & 0.434  \\
                 & F1280W &0.014 & 0.63 & 23 & 0.2 &  &  \\
                 & F1800W & 0.951 & 0.50 &9 & 0.2 &  &   \\
                 & F2550W & 0.132 & 0.79 &  5 & 0.1 &  &  \\
        source 2 & F560W &  0.177 & 1.00 & -   &  -   & 1.503 & 0.402 \\
                 & F1280W & 0.322 & 1.00 &  -  &  -   &       &      \\
                 & F1800W &  0.264 & 1.00 &  - &  -   &       &          \\
                 & F2550W & 0.264 & 1.00 &  -  &  -   &       &             \\
        source 3 & F560W &  0.225 & 1.00 &  -  &  -   & 2.047 & 0.649 \\
                 & F1280W & 0.001 & 0.60 &   -  &   -    &      &       \\
                 & F1800W & 0.546 & 3.69 &  -    &    -  &      &       \\
                 & F2550W & 0.003 & 1.00 &  -    &   -    &      &        \\

    \hline
    GRAL1131   & F560W  & 0.43                & 4.0            & 3 & 2 & $-$0.044 & 0.355  \\
            & F1280W & 0.44                & 1.3            & 3 & 2 \\
            & F1800W & 0.39                & 0.5            & - & -\\ 
            & F2100W & 0.38                & 0.5            & - & -\\
    \hline
    2M1134 & F560W & 0.073 & 2.89&   &   &  0.276 & $-$0.805  \\
    
    \hline
    J1251    &F560W  &0.11  &0.9 &- &-&0.740 &$-$0.340\\
            & F1280W &0.12  &0.5 &-  & - &      &\\
            & F1800W &0.097 &0.5 &-  & - &\\
            & F2100W &0.12  &0.5 &-  & - &\\
    \hline

       H1413 & F560W & 0.088 & 4.00 & - &      &0.647 & 0.311 \\
    \hline
    J2017 & F560W& 0.13 & 1.0 &- &-   & 0.235 & $-$0.390 \\
          & F1280W& 0.11 & 1.0 &5 &4.2 & &  \\
          & F1800W& 0.15 & 1.0 &- &-   &  &  \\
    \hline

   \end{tabular}
    \caption{Best fitting source parameters with positions given in the coordinate system of Table \ref{tab:data_summary}. The source centroids are restricted to be the same in all filters. Lens J2145 and filters which are not listed did not contain a detected lensed quasar host galaxy.}
    \label{tab:source}
\end{table*}

\addtocounter{table}{-1}

\begin{table*}
    \centering
    \begin{tabular}{llll|ll|lll}
    \hline
    \hline
    Lens    &Filter & r$_{\rm{s}}$ (arcsec)& n$_{\rm{s}}$ & n$_{\rm{max}}$ &  $\beta$ (arcsec) & dRa & dDec\\
    \hline
   WFI2033 &F560W  &0.01 & 2.7 &3 &2. & 1.210 & $-$1.377 \\
           &F1280W &0.02 & 0.8 &3 &2. &  \\
           &F1800W &0.23 & 0.5 & - & -  \\
    \hline     
    B2045 &F560W &0.11 & 2.15 &- &- & 0.596 & -2.391     \\
    \hline
    J2107 &F560W &0.02 & 0.97 & 9 & 0.55 & $-$1.857 & $-$0.943 \\
          &F1280W &0.01 & 3.98 & 7 & 2.00 &  & \\ 
          &F1800W &0.02 & 1.42 & 5 & 2.00 &  & \\ 
          &F2550W &0.02 & 0.50 & 5 & 1.02 &  & \\ 
    \hline
    J2205 & F560W & 0.110 & 1.21 &   -    &  -   &$-$1.020 & $-$0.033  \\
    \hline
  
    J2344 & F560W & 0.238 & 1.63 &  - &  -  & 0.393 & $-$0.341 \\
          & F1280W  & 0.287 & 0.50 & -  &  -  & \\
          & F1800W & 0.326 & 0.50 &  - &  -  & \\
          & F2100W & 0.324 & 0.50 & -  & -   & \\
    
    \end{tabular}
    \caption{Continued.}
    \label{tab:source}
\end{table*}

%% file: data_tables/lens_light_parameters.tex
\begin{table*}
    \centering
    \begin{tabular}{llllllll}
    \hline
    \hline
    Lens    &Filter & R$_{\rm{s}}$ (arcsec) & $q$ & $\theta$ (degrees) &dRA & dDec & R$_{\rm{s, sat}}$ \\
    \hline
    PSJ0147    &F560W  &  0.51 & 0.75 & 342 & 0.988 & 0.681 \\
    \hline
    J0248  &F560W  &0.10  & 0.22 & 327  & -0.359 & 0.831 &      \\ 
    \hline
     J0259  & F560W & 0.08 & 0.39 & 311  & 0.028 & 0.687 &  \\
    \hline
    MG0414  &F560W & 0.26 & 0.33 & 68  & -1.114 & 0.671 & \\
    \hline
    
    HE0435 & F560W   & 0.74 & 0.74 & 361  & -1.157 & -0.571 &  \\
    \hline
    J0457   &F560W & 1.56 & 0.74 & 307 & 2.225 & -1.897 \\
           &F1280W &0.10 & 0.62 & 303 & 2.225 & -1.897 \\
    \hline
    J0803 & F560W & 1.41 & 0.63 & 83  & -0.446 & 0.106 & \\
    \hline

    J0924   & F560W  & 0.26 & 0.80 & 355  & 0.357 & -0.799 & \\
            & F1280W  & 0.01 & 0.41 & 54  &          &       &  \\

    \hline
    HE1113 &F560W &0.94 & 0.91 & 266 & -0.310 & 0.262 \\
    \hline
    PG1115 & F560W & 0.96 & 0.57 & 126  & 0.770 & 0.588   \\
    \hline
    RXJ1131 &F560W & 1.29 & 0.75 & 300 & -2.008 & 0.587 \\
    \hline
    GRAL1131 & F560W &0.43 &0.60 & 173 &-0.073 &0.774\\ 
          & F1280W&0.072&0.40 &165 \\
    \hline 
    2M1134 & F560W & 0.94 & 0.59 & 237  & 1.196 & 1.555 & \\
    \hline
    J1251 & F560W  &0.38 &0.67 &282 &1.066 &-0.332 \\
         & F1280W &0.71 &0.50 & 298 \\
         & F1800W &0.077&0.41 & 218 \\    
    \hline     
    J2017 &F560W &1.57 & 0.48 & 336 & 0.436 & -0.109 \\
    \hline
    WFI2033 &F560W  &2.19 & 0.86 & 22  & 0.752 & -0.943 & 0.11\\
            &F1280W &2.20 & 0.70 & 347 &       &        & 0.10\\
    \hline
    B2045 &F560W  & 0.36 & 1 & -& 1.108 & -0.804 \\
         &F1280W  &0.45 & 1 & - &  \\
         &F1800W  &0.07 & 1 & - &  \\
         &F2550W  &0.59 & 1 & - &  \\
    \hline
    J2107 &F560W &0.68 & 0.52 & 241 & -1.947 & -1.456 \\
    \hline
   
    J2205 & F560W  & 0.80 & 0.50 & 23  & -0.906 & -0.131 \\
          & F1280W & 0.23 & 0.37 & 328  &       &      \\
    \hline
    
    J2344 & F560W   & 0.52 & 0.73 & 58  & 0.428 & -0.166 \\
           & F1280W & 0.69 & 0.64 & 29  &         &      \\

    \end{tabular}
    \caption{Best fitting deflector and satellite light parameters with positions given relative to the coordinate system in Table \ref{tab:data_summary}. Deflector light centroids are constrained to be the same in all filters. S\'ersic indices are held fixed to 4 for the deflector and satellite light. Luminous satellite positions are held fixed to the mass centroid given in Table \ref{tab:macromodel}. Lenses not included in this table did not have detected deflectors or satellites. \label{tab:light}}
\end{table*}

%% file: data_tables/psf_pars.tex
\begin{table*}
    \centering
    \begin{tabular}{llllll}
    \hline
    \hline
    Lens    &Filter & {\tt{jitter\_sigma}} & T (K) &  f$^a$ \\
    \hline
    PSJ0147 &F560W  &0.060 &1250 &0.38\\
         &F1280W &0.065 &467 \\
         &F1800W &0.072 &361 \\
         &F2550W &0.083 &291 \\
         \hline
    J0248 &F560W &0.061 & 1400 & 0.51 \\
         &F1280W &0.061 & 730 \\
         &F1800W &0.061 & 410  \\
         &F2550W &0.049 & 420   \\

         \hline
    J0259 & F560W & 0.059 & 1400 & 0.39 \\
         & F1280W &0.060 & 800  \\
         & F1800W &0.070 & 1400  \\
         & F2550W &0.07 & 2600  \\
         \hline
    MG0414 & F560W & 0.063 & 1100 & 0.19 \\
         & F1280W &0.062 & 720  \\
         & F1800W &0.066 & 420  \\
         & F2550W &0.066 & 540  \\
         \hline     
    HE0435 & F560W & 0.061 & 1100 & 0.5 \\
         & F1280W &0.062 & 670  \\
         & F1800W &0.068 & 650  \\
         & F2550W &0.079 & 370  \\
         \hline  
    J0457 & F560w & 0.068 & 696 & 0.26 \\
    & F1280w &0.062 & 922  \\
    & F1800w &0.065 & 683  \\
    & F2550w &0.064 & 721  \\
    \hline

    \hline     
    J0803 & F560W & 0.062 & 1700 & 0.36 \\
        & F1280W &0.064 & 920  \\
        & F1800W &0.068 & 660  \\
        & F2550W &0.080 & 1000  \\    
        \hline 
    J0924 & F560W & 0.063 & 1160 & 0.44 \\
        & F1280W &0.065 & 550  \\
        & F1800W &0.083 & 1000  \\
        & F2100W &0.097 & 450  \\    
        \hline     
    HE1113 & F560W & 0.068 & 500 & 0.31 \\
        & F1280W &0.062 & 420  \\
        & F1800W &0.069 & 320  \\
        & F2550W &0.072 & 920  \\    
        \hline   
    PG1115 & F560W & 0.059 & 1800 & 0.34 \\
        & F1280W &0.059 & 480  \\
        & F1800W &0.063 & 310  \\
        & F2550W &0.057 & 240  \\    
        \hline       
    GRAL1131 & F560W & 0.054 & 710 & 0.27 \\
             & F1280W & 0.062 & 470 \\
             & F1800W & 0.065 & 260 \\
             & F2550W & 0.079 & 480 \\

    \end{tabular}
    \caption{Best fitting PSF parameters. a) Fractional weighting of the zero and second fits extensions in the PSF model. 
    Set to 1 for all but F560W. }
    
    \label{tab:psf}
\end{table*}

\addtocounter{table}{-1}

\begin{table*}
    \centering
    \begin{tabular}{llllll}
    \hline
    \hline
    Lens    &Filter & {\tt{jitter\_sigma}} & T (K) &  f$^a$ \\
    \hline
    \hline
    2M1134 & F560W & 0.055 & 1500 & 0.74 \\
        & F1280W &0.060 & 780  \\
        & F1800W &0.064 & 600  \\
        & F2550W &0.074 & 600  \\
        \hline
    J1251 & F560W & 0.054 &420 & 0.35 \\
          & F1280W & 0.046 & 320 & \\
          &F1800W & 0.059 & 510 \\
          & F2550 & 0.086 & 710 \\
\hline
    
    H1413 & F560W & 0.054 & 1600 & 0.20 \\
        & F1280W &0.062 & 710  \\
        & F1800W &0.066 & 440  \\
        & F2550W &0.073 & 420  \\   
    \hline

    J2017 & F560W & 0.060 & 1100 & 0.34 \\
        & F1280W &0.063 & 1200  \\
        & F1800W &0.096 & 1400  \\
        & F2550W &0.113 & 1900  \\

\hline
    WFI2033 & F560W & 0.048 & 1200 & 0.029 \\
            & F1280W & 0.054 & 620 \\
            & F1800W & 0.062 & 440 \\
            & F2550W & 0.092 & 950 \\

\hline
    B2045 & F560W & 0.052 & 1100 & 0.19 \\
          & F1280W & 0.066 & 900 &   \\
          & F1800W & 0.071 & 1000 & \\
          & F2550W & 0.077 & 250 & \\

    \hline 
    J2107 & F560W & 0.051 &  1500 & 0.36 \\
          & F1280W & 0.057 & 570 \\
          & F1800W & 0.060 & 400 \\
          & F2550  & 0.056 & 450 \\

    \hline
    J2145 & F560W & 0.061 & 800 & 0.32 \\
         & F1280W &0.065 & 400  \\
        & F1800W &0.076 & 370  \\
        & F2550W &0.063 & 230  \\
    \hline    
   
    J2205 & F560W & 0.061 & 7400  & 0.6 \\
          & F1280W & 0.062 & 910  &  \\
          & F1800W & 0.075 & 1100 & \\
          & F2550  & 0.076 & 310 & \\

\hline

    \hline    
    J2344 & F560W & 0.060 & 570 & 0.44 \\
    & F1280W &0.060 & 560  \\
    & F1800W &0.047 & 270  \\
    & F2100W &0.084 & 340  \\
        \end{tabular}
    \caption{Continued.}
\end{table*}

%% file: data_tables/fluxes_table.tex
\begin{table*}
    \centering
    \begin{tabular}{lllllll}
    \hline
    \hline
    Lens           &image &  F560W &  F1280W & F1800W& F2100W & F2550W   \\
    \hline
    
    PSJ0147 & A & 0.503  & 2.09  & 2.913 & & 3.931 \\ 
            & B & 2.963  & 12.55  & 17.49 & & 23.42 \\ 
            & C & 8.019  & 35.89  & 49.75 & & 65.23 \\ 
            & D & 4.738  & 20.95  & 29.55 & & 40.00 \\

    \hline

    J0248    & A & 0.2054 & 1.272 & 1.981 & &2.66 \\
        & B & 0.208 & 1.285 & 2.011 & & 2.749 \\
        & C & 0.2559 & 1.635 & 2.57 & &3.429 \\
        & D   & 0.1968 & 1.31 & 2.059 & & 2.82 \\

    \hline
   
    J0259 & A & 0.1952  & 0.8245  & 1.219 &  & 1.589 \\ 
          & B & 0.2112  & 0.8443  & 1.258  & & 1.622 \\ 
          & C & 0.2102  & 0.8289  & 1.242  & & 1.58 \\ 
           & D & 0.3159  & 1.213  & 1.822  &  &2.332 \\ 

  \hline

    MG0414 & A & 3.734  & 17.57  & 23.38 & & 26.77 \\ 
           & B & 3.349  & 15.59  & 20.08 & & 23.89 \\ 
           & C & 1.366  & 6.412  & 8.547 & & 9.772 \\ 
           & D & 0.5579  & 2.548  & 3.366 &  & 3.856 \\ 

    \hline
  
    HE0435 & A & 0.6211  & 1.867  & 2.386  & & 3.731 \\ 
            & B & 0.613  & 1.865  & 2.392  & & 3.75 \\ 
            & C & 0.5305  & 1.676  & 2.179 & & 3.421 \\ 
            & D & 0.3686  & 1.107  & 1.395 & & 2.193 \\

    \hline
   
    J0457 &A &0.6359 &3.972 &5.880& &7.451\\
          &B &0.5363 &3.158 &4.688& &5.937\\
          &C &0.5028 &2.974 &4.428& &5.627\\
    \hline    
   
    J0803  & A & 0.5854  & 2.674  & 3.877 & & 4.811 \\ 
           & B & 0.4614  & 2.158  & 3.188 & & 3.896 \\ 
           & C & 0.1696  & 0.71  & 1.005 & & 1.253 \\ 
           & D & 0.3449  & 1.506  & 2.201 &  & 2.692 \\ 
    \hline    
    
   J0924 & A & 0.2893  & 0.9915  & 1.474  & 1.795 \\ 
         & B & 0.1238  & 0.4023  & 0.6111  & 0.7534 \\ 
         & C & 0.1066  & 0.4005  & 0.5983  & 0.7327 \\ 
         & D & 0.02076  & 0.0786  & 0.1632  & 0.2381 \\ 
           
    \hline     

HE1113  &A &1.091 &3.178 &4.952& &9.235 \\
        &B &1.227 &3.823 &5.942& &11.07\\
        &C &0.5267 &1.596 &2.535& &4.725\\
        &D &0.7292 &2.209 &3.470& &6.492\\
    \hline

PG1115    & A & 0.3929  & 1.562  & 2.092 &  & 3.311 \\ 
          & B & 0.4721  & 2.091  & 2.804 &  & 4.417 \\ 
          & C & 1.851  & 7.149  & 9.697 & & 15.7 \\ 
          & D & 1.95  & 7.773  & 10.54 & & 17.73 \\

    \hline

    RXJ1131 & A & 3.249  & 8.505  & 14.56 &  & 21.88 \\ 
            & B & 1.738  & 4.748  & 8.302 & & 12.14 \\ 
            & C & 1.072  & 3.071  & 5.957 & & 9.779 \\ 
            & D & 0.1856  & 0.473  & 0.7976 &  & 1.205 \\            
    \hline
    
    GRAL1131 & A & 0.1827  & 0.5883  & 1.007 &  & 1.78 \\ 
            & B & 0.2965  & 0.9  & 1.526  & & 2.78 \\ 
            & C & 0.6977  & 2.293  & 3.976 &  & 7.226 \\ 
            & D & 0.6831  & 2.269  & 3.956 & & 7.217 \\ 
    \hline        

2M1134       & A & 0.7718  & 6.154  & 8.363 & & 9.994 \\ 
            & B & 0.2113  & 1.513  & 2.048  & &2.455 \\ 
            & C & 0.8423  & 6.479  & 8.806  & &10.52 \\ 
            & D & 0.5453  & 3.4  & 4.621  &  &5.477 \\ 

    \hline 

    \end{tabular}
    \caption{Measured image fluxes in units of mJy}
    \label{tab:fluxes}
\end{table*}

\addtocounter{table}{-1}

\begin{table*}
    \centering
    \begin{tabular}{lllllll}
    \hline
    \hline
    Lens           &image &  F560W &  F1280W & F1800W& F2100W & F2550W   \\
    \hline   
    
    J1251 &
    A & 0.03881 & 0.09062 & 0.1183 & &0.162 \\ 
    & B & 0.07306 & 0.1857 & 0.2608 & &0.3415  \\
    & C & 0.158 & 0.4002 & 0.5988 &  &0.7457 \\
    & D & 0.1065 & 0.2427 & 0.3447 & & 0.429 \\
    \hline 
H1413     & A & 1.656  & 9.426  & 14.04 &  & 20.06 \\ 
          & B & 1.42  & 8.215  & 12.15 & & 17.49 \\ 
          & C & 1.291  & 7.928  & 11.78 & & 16.94 \\ 
          & D & 0.7111  & 4.01  & 5.84 & & 8.541 \\ 

    \hline

J2017 &A &0.2086 & 0.5754 &0.7729 & &  1.263\\
         &B &0.4004 &1.246 &1.557 & &2.626\\
         &C &0.3063 &0.8914 &1.126  & &1.923\\
         &D &0.1512 &0.5786 &0.733 & &1.348\\

    \hline

   \hline
   WFI2033  & A1 & 0.706  & 2.271  & 3.245 & & 5.072 \\ 
            & A2 & 0.494  & 1.572  & 2.285 & & 3.64 \\ 
            & C & 0.3649  & 1.162  & 1.659 & & 2.667 \\ 
            & D & 0.4485  & 1.372  & 1.936 & & 3.089 \\ 
         
    \hline

   B2045 &A &0.1574 &0.6142 &0.8243 & &1.315  \\
         &B &0.0090 &0.3598 &0.4849 & &0.6996 \\
         &C &0.1082 &0.4098  &0.5534 & &0.9343 \\
         &D &0.002711 &<0.001 &<0.002& &<0.003 \\

    \hline
J2107 & A & 1.700 & 12.410 & 18.434 &   &22.789\\
      & B & 1.136 & 11.83 & 20.71 &  &30.39 \\

\hline    
    J2145 & A & 1.551  & 4.114  & 5.683 &  & 9.768 \\ 
    & B & 0.9293  & 2.403  & 3.312 & & 5.44 \\ 
    & C & 4.542  & 12.18  & 16.73 & & 30.45 \\ 
    & D & 6.747  & 17.17  & 22.82 & & 40.31 \\ 
    
    \hline

   J2205    & A & 0.05613  & 0.2842  & 0.404 &   & 0.6366 \\ 
            & B & 0.08512  & 0.4373  & 0.6235&  & 0.9813 \\ 
            & C & 0.1576   & 0.7581  & 1.091 &  & 1.723 \\ 
            & D & 0.09296  & 0.4707  & 0.6667&  & 1.066 \\ 

    \hline 
 
    J2344     & A & 0.05161  & 0.1953  & 0.3284  & 0.4561 \\ 
          & B & 0.07012  & 0.2282  & 0.3929  & 0.5395 \\ 
          & C & 0.05368  & 0.1758  & 0.3  & 0.4139 \\ 
          & D & 0.07617  & 0.2825  & 0.4878  & 0.6846 \\

    \hline
    \end{tabular}
    \caption{Continued.}
\end{table*}